\documentclass{siamart0516}
\usepackage{latexsym,amssymb,enumerate,amsmath} 
  \usepackage{graphicx} 
  \usepackage{tikz}
  \usepackage[square,sort,comma,numbers]{natbib}
  \usepackage{hyperref}

\numberwithin{equation}{section}

\usepackage{latexsym,amssymb,enumerate,amsmath} 
  \usepackage{graphicx} 
  \usepackage{tikz}
  \usepackage[square,sort,comma,numbers]{natbib}
  \usepackage{hyperref}

\usepackage{mathtools} 
  \usepackage{graphicx} 
\usepackage{soul}
\usepackage{halloweenmath}

\newcommand{\dd}{\text{d}}
\def\eps{\varepsilon}
\def\E{\mathbb{E}}
\def\P{\mathbb{P}}
\def\R{\mathbb{R}}

\def\N{\mathbb{N}}

\usepackage{amsfonts}
\usepackage{graphicx}
\usepackage{epstopdf}
\usepackage{algorithmic}
\ifpdf
  \DeclareGraphicsExtensions{.eps,.pdf,.png,.jpg}
\else
  \DeclareGraphicsExtensions{.eps}
\fi

\newcommand{\TheTitle}{Boundary homogenization for partially reactive patches}
\newcommand{\ShortTitle}{Boundary homogenization for partially reactive patches}
\newcommand{\TheAuthors}{Claire E. Plunkett and Sean D. Lawley}

\headers{\ShortTitle}{\TheAuthors}

\title{{\TheTitle}\thanks{%Submitted to the editors \today.
\funding{This work was supported by the National Science Foundation (CAREER DMS-1944574, DMS-1814832, and DMS-2325258).}}}

\author{Claire E. Plunkett\thanks{Department of Mathematics, University of Utah, Salt Lake City, UT 84112 USA.}\and {Sean D. Lawley\thanks{Department of Mathematics, University of Utah, Salt Lake City, UT 84112 USA (\texttt{lawley@math.utah.edu}).}}}
\date{\today}

\ifpdf
\hypersetup{
  pdftitle={\TheTitle},
  pdfauthor={\TheAuthors}
}
\fi

\begin{document}

\maketitle

%%%%%%%%%%%%%%%%%%%%%%%%%%%%%%%%%%%%%%%%%%%%%%%%%%%%%%%%%%%%%%%%%%%%%%%%%%%%%%%%%%%%%%%%%%%%%%%%%%%%%%%%%%%%%%%%%%%%%%%%%%%%%%%%%%%%%%%%%%%%%%%%%%%%%%%%%%%%%%%%%%%%%%%%%%%%%%%%%%%%%%%%%%%%%%%%%%%%%%%%%%%%%%%%%%%%%%%%%%%%%%%%%%%%%%%%%%%%%%%%%%%%%%%%%%%%%%%%%%%%%%%%%%%%%%%%%%%%%%%%%%%%%%%%%%%%%%%%%%%%%%%%%%%%%%%%%%%%%%%%

\begin{abstract}
A wide variety of physical, chemical, and biological processes involve diffusive particles interacting with surfaces containing reactive patches. The theory of boundary homogenization seeks to encapsulate the effective reactivity of such a patchy surface by a single trapping rate parameter. In this paper, we derive the trapping rate for partially reactive patches occupying a small fraction of a surface. We use matched asymptotic analysis, double perturbation expansions, and homogenization theory to derive formulas for the trapping rate in terms of the far-field behavior of solutions to certain partial differential equations (PDEs). We then develop kinetic Monte Carlo (KMC) algorithms to rapidly compute these far-field behaviors. These KMC algorithms depend on probabilistic representations of PDE solutions, including using the theory of Brownian local time. We confirm our results by comparing to KMC simulations of the full stochastic system. We further compare our results to prior heuristic approximations.
\end{abstract}

% REQUIRED
% REQUIRED

%\tableofcontents
%%%%%%%%%%%%%%%%%%%%%%%%%%%%%%%%%%%%%%%%%%%%%%%%%%%%%%%%%%%%%%%%%%%%%%%%%%%%%%%%%%%%%%%%%%%%%%%%%%%%%%%%%%%%%%%%%%%%%%%%%%%%%%%%%%%%%%%%%%%%
\section{Introduction}

Many biological, chemical, and physical processes involve diffusive particles being trapped at localized surface patches or transported through small pores. Examples include proteins binding to receptors on a cell membrane \cite{berg-physics-1977, roberts-role-2014}, industrial processes such as filtration \cite{gahn-singular-2021, saxena-membrane-based-2009} and gas separation \cite{baker-gas-2014}, water transpiration through plant stomata \cite{wolf-optimal-2016}, reactions on porous catalyst support structures \cite{keil-diffusion-1999}, nanopore sensing for detection of biomolecules \cite{qiao-efficient-2021}, biological microdevices composed of microchambers separated by cell layers on porous membranes \cite{gahn-singular-2021, huh-reconstituting-2010}, and transport through the nuclear boundary via the nuclear pore complex \cite{hoogenboom-physics-2021}.

Mathematical models of such processes frequently use the diffusion equation with mixed Dirichlet-Neumann boundary conditions, where patches of the boundary are perfectly reactive (using Dirichlet boundary conditions) and the remainder is perfectly reflective (using Neumann boundary conditions). As a prototypical scenario, if $u=u(x,y,z,t)\ge0$ denotes the concentration of particles at a distance $z\ge0$ above a planar surface, then $u$ satisfies the diffusion equation,
\begin{align*}%\label{upde}
\partial_{t}u
=D\Delta u,\quad z>0,\,(x,y)\in\R^{2},
\end{align*}
where $D>0$ is the diffusivity and $\Delta$ denotes the Laplacian acting on $(x,y,z)$, with mixed boundary conditions at the surface,
\begin{align}
\begin{split}\label{mixed}
u
&=0,\quad z=0,\,\text{$(x,y)$ in a reactive patch},\\
\partial_{z}u
&=0,\quad z=0,\,\text{$(x,y)$ not in a reactive patch}.
\end{split}
\end{align}
However, the mixed boundary conditions \eqref{mixed} can be challenging to work with and are often replaced by a single homogeneous Robin boundary condition of the form
\begin{align}\label{homog}
D\partial_{z}
u = \overline{\kappa} u,\quad z=0,\,(x,y)\in\R^{2},
\end{align}
using a technique called ``boundary homogenization" \cite{berezhkovskii-boundary-2004, zwanzig-diffusion-controlled-1990}. The effective ``trapping rate" parameter $\overline{\kappa} > 0$ in \eqref{homog} encompasses the effective trapping properties of the ``patchy'' surface. In the case of roughly evenly distributed circular patches of radius $a>0$ occupying a small fraction $\sigma\ll1$ of the surface, the trapping rate is given by \cite{berg-physics-1977, shoup1982}
\begin{align}\label{bp}
\overline{\kappa}_{0}
=\frac{4D\sigma}{\pi a}.
\end{align}
More generally, trapping rates have been derived in a great variety of scenarios, including particles diffusing above patchy planes \cite{muratov-boundary-2008, bernoff-boundary-2018, berezhkovskii-homogenization-2006, belyaev-effective-1999, berezhkovskii-boundary-2004, makhnovskii-trapping-2006}, patchy particles diffusing above patchy planes \cite{plunkett-boundary-2023}, bimolecular reactions between two diffusive particles where either one is patchy \cite{zwanzig-diffusion-controlled-1990, dagdug-boundary-2016, lindsay-first-2017, lawley-how-2019} or both are patchy \cite{plunkett-bimolecular-2021}, and a patchy particle diffusing above an entirely reactive plane \cite{lawley-boundary-2019}. 

\begin{figure}%[h]
\centering
\includegraphics[width=0.6\textwidth]{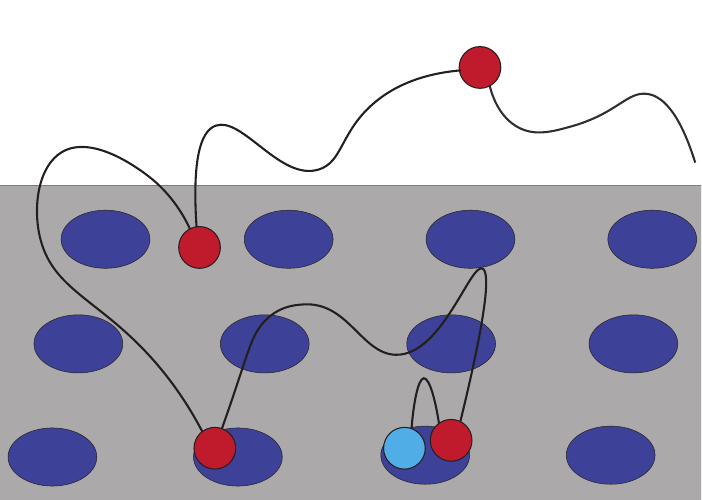}
\caption{A diffusing particle first reflecting off of the reflective portion of the plane (gray region) and then off of partially reactive patches (dark blue disks) twice before being absorbed at a partially reactive patch. Absorption is indicated by the red particle becoming light blue.}
\label{fig_schem_robin_example}
\end{figure}

The majority of prior work in boundary homogenization assumes the patches are perfectly reactive, which corresponds to the mixed Dirichlet-Neumann boundary conditions in \eqref{mixed}. However, another option is \textit{partially reactive} patches, which is described by mixed Robin-Neumann boundary conditions of the form
\begin{align}
\begin{split}\label{mixed2}
D\partial_{z}u
&=\kappa_{\text{p}}u,\,\qquad z=0,\,\text{$(x,y)$ in a reactive patch},\\
\partial_{z}u
& =0,\qquad\quad z=0,\,\text{$(x,y)$ not in a reactive patch}.
\end{split}
\end{align}
Here, $\kappa_{\text{p}}>0$ denotes the reactivity of each patch (notice that \eqref{mixed2} reduces to \eqref{mixed} in the limit $\kappa_{\text{p}}\to\infty$). Figure~\ref{fig_schem_robin_example} illustrates a particle diffusing above a surface with mixed Robin-Neumann boundary conditions \eqref{mixed2} reflecting several times before being absorbed at one of the partially reactive patches. Prior work 
\cite{zwanzig-time-1991, berezhkovskii-boundary-2004} has posited the following heuristic formula to take a trapping rate $\overline{\kappa}_{0}$  which homogenizes the Dirichlet-Neumann conditions in \eqref{mixed} and obtain a trapping rate $\overline{\kappa}_{\text{heur}}$ which homogenizes the Robin-Neumann conditions in \eqref{mixed2},
\begin{align}\label{interp}
\overline{\kappa}_{\text{heur}}
=\frac{\sigma\kappa_{\text{p}}\overline{\kappa}_{0}}{\sigma\kappa_{\text{p}}+\overline{\kappa}_{0}},
\end{align}
where $\sigma\in(0,1)$ denotes the fraction of the surface covered by patches. That is, it has been claimed \cite{zwanzig-time-1991, berezhkovskii-boundary-2004} that if the mixed Dirichlet-Neumann boundary conditions in \eqref{mixed} are well-approximated by the homogeneous Robin boundary condition in \eqref{homog} with $\overline{\kappa}=\overline{\kappa}_{0}$, then the mixed Robin-Neumann boundary conditions in \eqref{mixed2} are well-approximated by the homogeneous Robin boundary condition in \eqref{homog} with $\overline{\kappa}=\overline{\kappa}_{\textup{heur}}$.

%%%%%%%%%%%%%%%%%%%%%%%%%%%%%%%%%%%%%%%
%%%%%%%%%%%%%%%%%%%%%%%%%%%%%%%%%%%%%%%

Numerous biophysical and chemical phenomena involve partially reactive patches, such as reactions involving an energetic activation or entropic barrier, stochastic gating, target recognition, permeation, and inclusions in liquid crystals \cite{lindenberg-imperfect-2019, lawley-new-2015}. Additionally, other modeling needs or techniques may support using partially reactive patches. These include models where a single partially reactive patch replaces clusters of reactive disks \cite{berezhkovskii-trapping-2013, berezhkovskii-communication-2012}, or how partially reactive boundary conditions allow the diffusing particle to start on the target without immediately reacting \cite{lindenberg-imperfect-2019}. Some recent developments in modeling and analyzing partially reactive boundaries in terms of Brownian local time include \cite{lindenberg-imperfect-2019, grebenkov-spectral-2019, grebenkov-diffusion-2020, schumm-search-2021, bressloff-diffusion-mediated-2022, bressloff-narrow-2022, chaigneau-first-passage-2022}.

\begin{figure}%[h]
\centering
\includegraphics[width=\textwidth]{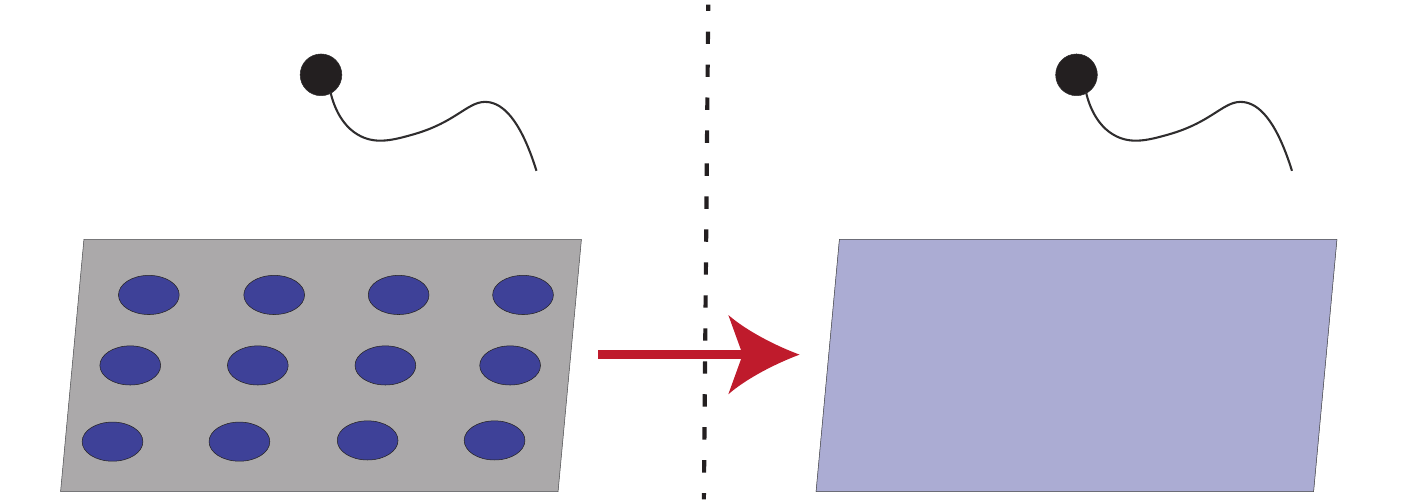}
\caption{Homogenization of a patchy surface. The left side illustrates a particle diffusing above the plane with blue partially reactive patches where the gray plane is otherwise reflective. The right side has a lighter blue plane to indicate that it is partially reactive everywhere, with a reactivity parameter less than the reactivity of the blue patches on the left.}
\label{fig_schem_homogenization}
\end{figure}

In this paper, we perform boundary homogenization for a plane with small partially reactive patches (see Figure~\ref{fig_schem_homogenization} for an illustration). Specifically, we derive the homogenized trapping rate as in \eqref{homog} for the mixed Robin-Neumann boundary conditions in \eqref{mixed2}. Assuming the patches are well-separated disks of radius $a>0$, we derive the following effective trapping rate $\overline{\kappa}$ in the limit that the patches occupy a small fraction $\sigma\ll1$ of the surface,
\begin{align}\label{main}
\overline{\kappa}
=
\begin{dcases*}
\frac{4D\sigma}{\pi a} & \text{if }$a\kappa_{\text{p}}/D\gg1$,\\
\frac{4D\sigma}{\pi a}\frac{c_{0}(a\kappa_{\text{p}}/D)}{2/\pi} & \text{if }$a\kappa_{\text{p}}/D=O(1)$,\\
\frac{4D\sigma}{\pi a}\frac{a\kappa_{\text{p}}/D}{2/\pi}K & \text{if }$a\kappa_{\text{p}}/D\ll1$.
\end{dcases*}
\end{align}
In \eqref{main}, $c_{0}=c_{0}(a\kappa_{\text{p}}/D)$ is a dimensionless function of the ratio $a\kappa_{\text{p}}/D$ and is akin to the ``capacitance'' of a single partially reactive disk. Further, $K>0$ is a constant and is akin to the ``capacitance'' of a single disk with a fixed unit flux. Using a probabilistic analysis and computation using Brownian local time, we find that
\begin{align*}
K
\approx 0.{5854},
\end{align*}
and that $c_{0}(\kappa')$ is well-approximated by the sigmoidal function,
\begin{align*}
c_{0}(\kappa')
&\approx\frac{(2/\pi) \kappa'}{\kappa'+2/(\pi K)}.
\end{align*}
Indeed, we show that the following explicit formula agrees with the trapping rate $\overline{\kappa}$ in \eqref{main} to within 5\%,
\begin{align*}
\overline{\kappa}
\approx
\frac{4D\sigma}{\pi a }\frac{(2/\pi) a\kappa_{\text{p}}/D}{a\kappa_{\text{p}}/D+2/(\pi K)}.
\end{align*}

We obtain \eqref{main} by performing matched asymptotic analysis and homogenization, using single and double perturbation expansions in the patch radii and the patch reactivity. The quantities $c_{0}$ and $K$ arise via the far-field behavior of the solutions to certain partial differential equations (PDEs) describing the particle concentration near a patch (i.e.\ the ``inner solutions'' in the matched asymptotic analysis \cite{ward_strong_1993}). Building on the algorithms devised by Bernoff, Lindsay, and Schmidt \cite{bernoff-boundary-2018}, we develop kinetic Monte Carlo (KMC) algorithms to rapidly compute $c_{0}$ and $K$ to high accuracy. Our algorithms rely on probabilistic representations of the inner solutions (either the ``splitting probability'' of a single particle or the Brownian local time of a single particle). These two algorithms extend the classical calculation of the capacitance of a perfectly reactive disk (the so-called ``electrified disk problem'' dating back to Weber \cite{weber1873, sneddon1966, jackson-classical-1999}). Additionally, we develop a third KMC algorithm to simulate the entire stochastic system and confirm the trapping rate in \eqref{main}. 

The rest of the paper is organized as follows. In section~\ref{stoch}, we formulate the problem in terms of a stochastic process and an associated PDE. In section~\ref{math}, we apply matched asymptotic analysis to derive the trapping rate in \eqref{main}. In sections~\ref{kmc1}-\ref{kmc3}, we develop the three aforementioned {{KMC}} algorithms and compare the results of our mathematical analysis to stochastic simulations.\footnote{The associated code is available at \url{https://github.com/ceplunk/KMC_partial_reactivity}.} In section~\ref{comparison}, we compare our asymptotic trapping rate in \eqref{main} to the heuristic trapping rate in \eqref{interp}. We find that these two trapping rates generally agree, with a maximum relative difference around 17\%. We conclude with a brief discussion.

%%%%%%%%%%%%%%%%%%%%%%%%%%%%%%%%%%%%%%%
%%%%%%%%%%%%%%%%%%%%%%%%%%%%%%%%%%%%%%%

%%%%%%%%%%%%%%%%%%%%%%%%%%%%%%%%%%%%%%%%%%%%%%%%%%%%%%%%%%%%%%%%%%%%%%%%%%%%%%%%%%%%%%%%%%%%%%%%%%%%%%%%%%%%%%%%%%%%%%%%%%%%%%%%%%%%%%%%%%%%%%%%%%%%%%%%%%%%%%%%
\section{Stochastic problem formulation}\label{stoch}

Consider a Brownian particle with diffusivity $D>0$ diffusing in three dimensions above an infinite plane. Suppose the plane is reflecting except for an infinite collection of roughly evenly distributed, partially reactive disks with common radius $a>0$ and reactivity $\kappa_{\text{p}}>0$ centered at $\{(\tilde{x}_{n},\tilde{y}_{n})\}_{n=1}^{\infty}$. Assume the fraction of the surface covered in patches is given by 
\begin{align}\label{mr}
\sigma:=\lim_{l\to\infty}\frac{\pi a^{2}}{(2l)^{2}}\int_{[-l,l]^{2}}\sum_{m=1}^{\infty}\delta(\tilde{x}-\tilde{x}_{m})\delta(\tilde{y}-\tilde{y}_{m})\,\dd \tilde{x}\,\dd \tilde{y}
=\frac{\pi a^{2}}{L^{2}}\ll1,
\end{align}
and assume that the patches are well-separated in the sense that the patch radius is much less than the distance between any pair of patches,
\begin{align}\label{wellsep}
a
\ll\inf_{m\neq n}\sqrt{(\tilde{x}_{m}-\tilde{x}_{n})^{2}+(\tilde{y}_{m}-\tilde{y}_{n})^{2}}.
\end{align}
Notice that the integral in \eqref{mr} is simply the number of patch centers in the square $[-l,l]^{2}$. 

We non-dimensionalize the problem by rescaling time by $L^2/D$ and rescaling space by $L$ so that the particle has unit diffusivity, the patch centers are
\begin{align}
\begin{split}\label{rescaling}
\mathbf{x}_n
&= ({x_{n}}, {y_{n}}, 0 ) := (\tilde{x}_n/L, \tilde{y}_n/L, 0),
\end{split}
\end{align}
and the disks have dimensionless radius
\begin{align*}
\eps
&:= a/L \ll 1,
\end{align*}
with dimensionless reactivity
\begin{align*}
\kappa
&:= L \kappa_{\text{p}}/D.
\end{align*}
We denote the $n$th patch by
\begin{align*}
\partial \omega_{\eps,n} = \{ (x,y) \, : \, (x- {x_{n}})^2 + (y-{y_{n}})^2 < \eps^2 \},
\end{align*}
and the union of patches by $\partial \omega_\eps = \cup_{n=1}^\infty \partial \omega_{\eps,n}$. We denote the three-dimensional position of the particle at time $t \geq 0$ as
\begin{align*}
\mathbf{X}(t) = ( X(t), Y(t), Z(t)) \in \R^2 \times [0, \infty).
\end{align*}

In order to study the absorption time, we introduce the local time $\ell_{\eps}(t)$ of all patches \cite{grebenkov-paradigm-2020,grebenkov-probability-2019}, 
\begin{align*}
\ell_{\eps}(t) = \lim_{h \to 0} \frac{1}{2h} \int_0^t \mathbf{1}_{ \mathbf{X}(s) \in \omega_\eps^h}\, \dd s,
\end{align*}
where $\mathbf{1}_{A}$ denotes the indicator function on an event $A$ (i.e.\ $\mathbf{1}_{A}=1$ if $A$ occurs and $\mathbf{1}_{A}=0$ otherwise) and $\omega_\eps^h$ denotes the cylinders of height $h>0$ above all the patches,
\begin{align*}
\omega_\eps^h = \cup_{n=1}^\infty \{ (x,y,z) \, : \, (x- {x_{n}})^2 + (y-{y_{n}})^2 < \eps^2, 0 \leq z \leq h \}.
\end{align*}
We define a stopping time $\tau$ to indicate the first time that the particle is absorbed by one of the partially reactive patches,
\begin{align*}
\tau := \inf \left\{ t > 0 \, : \, \ell_{\eps}(t) > E/\kappa \right\},
\end{align*}
where $E$ is an independent exponential random variable with unit mean (so that $E/\kappa$ is an exponential random variable with rate $\kappa$) \cite{grebenkov-paradigm-2020}. It is well-known that this definition of $\tau$ yields a Robin boundary condition in the corresponding Kolmogorov equation (see \eqref{eq_s_all}) \cite{grebenkov-paradigm-2020}. The condition that $\ell_{\eps}(t)>E/\kappa$ for absorption before time $t$ means that the particle must spend some time on the patch before absorption, where smaller values of $\kappa$ correspond to requiring more time on the patch before absorption.

The probability distribution of $\tau$ is described by its survival probability,
\begin{align}\label{Sdef}
S(\mathbf{x},t) 
=S(x,y,z,t)
:= \P \left( \tau > t \, | \, \mathbf{X}(0) = \mathbf{x} \right),
\end{align}
where we have conditioned that the particle starts at position $\mathbf{x}=(x,y,z)$. This survival probability satisfies the following diffusion equation with mixed boundary conditions and unit initial condition \cite{pavliotis2016},
\begin{subequations}
\label{eq_s_all}
\begin{align}
\partial_t S &= \Delta S, \quad z>0,\,t > 0, \label{eq_s_diffeq}
\\
\partial_z S &= \kappa S, \quad z=0,\,(x,y) \in \partial \omega_\eps, \label{eq_s_bc1}
\\
\partial_z S &= 0, \quad z=0,\,(x,y) \notin \partial \omega_\eps, \label{eq_s_bc2}
\\
S &= 1, \quad z > 0, \, t = 0, \label{eq_s_ic}
\\
\lim_{z \to \infty} S &= 1. \label{eq_s_ff}
\end{align}
\end{subequations}

%%%%%%%%%%%%%%%%%%%%%%%%%%%%%%%%%%%%%%%%%%%%%%%%%%%%%%%%%%%%%%%%%%%%%%%%%%%%%%%%%%%%%%%%%%%%%%%%%%%%%%%%%%%%%%%%%%%%%%%%%%%%%%%%%%%%%%%%%%%%%%%%%%%%%%%%%%%%%%%%
\section{Matched asymptotic analysis}\label{math}

We now approximate the survival probability $S$ in \eqref{Sdef}-\eqref{eq_s_all} in the singular limit of vanishing patch radius ($\eps\to0$) using the method of matched asymptotic expansions. 

%%%%%%%%%%%%%%%%%%%%%%%%%%%%%%%%%%%%%%%%%%%%%%%%%%%%%%%%%%%%%%%%%%%%%%%%%%%%%%
\subsection{Outer expansion}

As $\eps \to 0$, the surface of the plane becomes perfectly reflecting, and thus
\begin{align}\label{obv1}
\lim_{\eps \to 0} S = 1.
\end{align}
We expect that $S$ has a 3-dimensional boundary layer in a neighborhood of each patch on the plane, so we introduce the outer double or single perturbation expansions for different cases of reactivity $\kappa$ magnitude. In particular, our final trapping rate formulas are valid in the cases that either $\kappa$ is small, large, or moderate (i.e.\ $\kappa=o(1)$, $\kappa^{-1}=o(1)$, or $\kappa\neq o(1)$ and $\kappa^{-1}\neq o(1)$). If $\kappa = o(1)$ or $\kappa^{-1} = o(1)$, we expand in patch radius $\eps$ and powers of reactivity $\kappa^{\gamma}$,
\begin{align}
S \sim 1 + \eps S_{1,0} + \kappa^{\gamma} S_{0,1} + \eps^2 S_{2,0} + \eps \kappa^{\gamma} S_{1,1} + \kappa^{2\gamma} S_{0,2} + \cdots, \label{eq_outer_expansion}
\end{align}
which is valid away from the boundary layers, where $\gamma \in \{ -1, 1 \}$ is chosen so that $\kappa^{\gamma} =  o(1) $. Plugging this outer expansion into \eqref{eq_s_all} shows that the functions $S_{i,j}$ must satisfy
\begin{align*}
\partial_t S_{i,j} &= \Delta S_{i,j}, \quad z>0,\,t > 0,
\\
\partial_z S_{i,j} &= 0, \quad z = 0, (x,y) \notin \cup_{n=1}^\infty ({x_{n}}, {y_{n}}), \, t > 0.
\end{align*}
Notice that from the perspective of the outer solution, the patches have shrunk to points $\cup_{n=1}^\infty ({x_{n}}, {y_{n}})$.

If neither $\kappa = o(1)$ nor $\kappa^{-1} = o(1)$, we can proceed with a single perturbation expansion in patch radius $\eps$,
\begin{align}
S \sim 1 + \eps S_{1} + \eps^2 S_{2} + \cdots, \label{eq_outer_expansion_single}
\end{align}
which is valid away from the boundary layers. Plugging this outer single expansion into \eqref{eq_s_all} shows that the functions $S_{i}$ must satisfy
\begin{align*}
\partial_t S_{i} &= \Delta S_{i}, \quad z>0,\,t > 0,
\\
\partial_z S_{i} &= 0, \quad z = 0, (x,y) \notin \cup_{n=1}^\infty ({x_{n}}, {y_{n}}), \, t > 0.
\end{align*}
We refer to this case as Case 0 and primarily focus on the double perturbation expansion cases.

%%%%%%%%%%%%%%%%%%%%%%%%%%%%%%%%%%%%%%%%%%%%%%%%%%%%%%%%%%%%%%%%%%%%%%%%%%%%%%
\subsection{Inner expansion} \label{sec_inner}

We now derive the inner expansion for the boundary layers surrounding each partially reactive patch in order to determine the singular behavior of the outer expansion as
\begin{align*}
\mathbf{x} \to ({x_{n}}, {y_{n}}, 0)\quad\text{for fixed }n\in\N.
\end{align*}
This will determine the behavior of the functions $S_{i,j}$ or $S_i$ as the initial location of the diffusing particle approaches a partially reactive patch.

Introduce the stretched coordinates
\begin{align*}
\mu &= \eps^{-1} (x - {x_{n}}),
\\
\nu &= \eps^{-1} (y - {y_{n}}),
\\
\eta &= \eps^{-1} z,
\end{align*}
and define the inner solution $w$ as a function of the stretched coordinates,
\begin{align*}
w(\mu, \nu, \eta, t) := S( \eps \mu + {x_{n}}, \eps \nu + {y_{n}}, \eps \eta , t).
\end{align*}
Above the $\eta = 0$ plane, $w$ must satisfy
\begin{align}
\eps^2 \partial_t S = \eps^2 \Delta S = \Delta w = \eps^2 \partial_t w. \label{eq_inner_bulk}
\end{align}
Using the boundary conditions \eqref{eq_s_bc1} and \eqref{eq_s_bc2} from the original problem, we derive the following boundary conditions for $w$ at $\eta = 0$, 
\begin{subequations}
\label{eq_inner}
\begin{align}
\partial_\eta w &= \eps \kappa w, \quad \eta = 0, \,\mu^2 + \nu^2 < 1, \label{eq_inner_bc1}
\\
\partial_\eta w &= 0, \quad \eta = 0, \,\mu^2 + \nu^2 > 1. \label{eq_inner_bc2}
\end{align}
\end{subequations}
We aim to expand $w$ as $\eps \to 0$. Since the product $\eps \kappa$ appears in the inner problem and we have not restricted the order of $\kappa$, we must use different inner expansions depending on the orders of $\eps \kappa$ and $\kappa$. We consider three separate cases for the inner expansions. A summary of these cases, including the behavior of $\eps \kappa$, $\kappa$, and the resulting {trapping rate}, is in Table \ref{table_sum}.

\begin{table}[h]
\centering
\caption{Cases for the behavior of the {trapping rate}.} \label{table_sum}
\begin{tabular}{|l|c|c|cc|}
\hline
& \textbf{Case I} & \textbf{Case II} & \multicolumn{2}{c|}{\textbf{Case III}} \\ \hline
Behavior of $\eps \kappa$ & $\eps \kappa \to \infty$ & $\eps \kappa \to C \in (0, \infty)$ & \multicolumn{2}{c|}{$\eps \kappa \to 0$} \\ \hline
Subcases & N/A & N/A & \multicolumn{1}{c|}{\begin{tabular}[c]{@{}c@{}}$\eps \kappa \propto \eps^\beta$ for\\ $\beta \in \N$\end{tabular}} & \begin{tabular}[c]{@{}c@{}}no $\beta \in \N$ such\\ that $\eps \kappa \propto \eps^\beta$\end{tabular} \\ \hline
Approach & \begin{tabular}[c]{@{}c@{}}Expansion in\\ $\eps$ and $\kappa^{-1}$\end{tabular} & \begin{tabular}[c]{@{}c@{}}Expansion in\\ $\eps$ and $\kappa^{-1}$\end{tabular} & \multicolumn{1}{c|}{\begin{tabular}[c]{@{}c@{}}Expansion\\ in $\eps$\end{tabular}} & \begin{tabular}[c]{@{}c@{}}Expansion in\\ $\eps$ and $\eps \kappa$\end{tabular} \\ \hline
Inner behavior & $w \sim 1 - \frac{2}{\pi \rho}$ & $w \sim 1 - \frac{c_0(\eps \kappa)}{\rho}$ & \multicolumn{2}{c|}{$w \sim 1 - \frac{\eps \kappa K}{\rho}$} \\ \hline
\begin{tabular}[c]{@{}l@{}}Trapping rate\end{tabular} & $4 \eps$ & $2 \pi \eps c_0(\eps \kappa)$ & \multicolumn{2}{c|}{$2 \pi \eps^2 \kappa K$} \\ \hline
\end{tabular}
\end{table}

%%%%%%%%%%%%%%%%%%%%%%%%%%%%%%%%%%%%%%%%%%%%%%%%%%%%%%%%%%%%%%%%%%%%%%%%%%%%%%%%%
\subsubsection{Case I: $\eps\kappa\to\infty$}

If $\eps \kappa \to \infty$ as $\eps \to 0$, then $\kappa \to \infty$ and $\kappa^{-1} \ll \eps$. We introduce the double perturbation expansion in $\eps \ll 1$ and $k := \kappa^{-1} \ll 1$,
\begin{align}
w = \sum_{i=0}^\infty \sum_{j=0}^\infty \eps^i k^j w_{i,j}. \label{eq_expansion_i}
\end{align}
We plug this expansion into the inner boundary condition on the disk \eqref{eq_inner},
\begin{align*}
\sum_{i=0}^\infty \sum_{j=0}^\infty \eps^i k^{j+1} \partial_\eta w_{i,j} = \sum_{i=0}^\infty \sum_{j=0}^\infty \eps^{i+1} k^{j} w_{i,j}, \quad \eta = 0, \,\mu^2 + \nu^2 < 1.
\end{align*}
We rearrange this equation to obtain the following condition on the disk,
\begin{align*}
\sum_{j=1}^\infty k^j \partial_\eta w_{0,j-1} - \sum_{i=1}^\infty \eps^i w_{i-1,0} + \sum_{i=1}^\infty \sum_{j=1}^\infty \eps^i k^j (\partial_\eta w_{i,j-1} - w_{i-1,j} ) = 0. 
\end{align*}
Since $k \ll \eps$, the leading order term in this boundary condition is
\begin{align*}
w_{0,0} = 0, \quad \eta = 0, \,\mu^2 + \nu^2 < 1,
\end{align*}
which describes a perfectly reactive unit disk. Since $w$ experiences a reflecting boundary condition away from the disk and satisfies $\Delta w = \eps^2 \partial_t w$ in the bulk, we have that $w_{0,0}$ satisfies
\begin{align*}
\Delta w_{0,0} &= 0, \quad \eta > 0, (\mu, \nu) \in \R^2,
\\
w_{0,0} &= 0, \quad \eta = 0, \,\mu^2 + \nu^2 < 1,
\\
\partial_\eta w_{0,0} &= 0, \quad \eta = 0, \,\mu^2 + \nu^2 > 1.
\end{align*}
It follows from the solution to the electrified disk problem that $w_{0,0}$ has far-field behavior \cite{weber1873, sneddon1966, jackson-classical-1999}
\begin{align}
w_{0,0} \sim \alpha \bigg( 1 - \frac{2}{\pi \rho} \bigg), \quad \text{ as } \rho:= \sqrt{\mu^2 + \nu^2 + \eta^2} \to \infty, \label{eq_far_field_i}
\end{align}
for some constant $\alpha$ which will be determined by matching to the outer solution.

%%%%%%%%%%%%%%%%%%%%%%%%%%%%%%%%%%%%%%%%%%%%%%%%%%%%%%%%%%%%%%%%%%%%%%%%%%%%%%%%%
\subsubsection{Case II: $\eps\kappa\to C\in(0,\infty)$}

If $\eps \kappa \to C \in (0, \infty)$ as $\eps \to 0$, then $\kappa \to \infty$ and that $\kappa^{-1} \sim \eps$. We use the same double perturbation expansion \eqref{eq_expansion_i} in $\eps \ll 1$ and $k := \kappa^{-1} \ll 1$ as in Case I. Using \eqref{eq_inner}, the leading order terms on the boundary at the disk are now
\begin{align*}
\partial_\eta w_{0,0} = \eps \kappa w_{0,0}, \quad \eta = 0, \,\mu^2 + \nu^2 < 1,
\end{align*}
which corresponds to a partially reactive boundary condition on the unit disk. We observe that $w_{0,0}$ satisfies
\begin{align*}
\Delta w_{0,0} &= 0, \quad \eta > 0, (\mu, \nu) \in \R^2,
\\
\partial_\eta w_{0,0} &= \eps \kappa w_{0,0}, \quad \eta = 0, \,\mu^2 + \nu^2 < 1,
\\
\partial_\eta w_{0,0} &= 0, \quad \eta = 0, \,\mu^2 + \nu^2 > 1.
\end{align*}
We show below that $w_{0,0}$ has far-field behavior
\begin{align}
w_{0,0} \sim \alpha \left( 1 - \frac{c_0(\eps \kappa)}{\rho} \right), \quad \text{ as } \rho:= \sqrt{\mu^2 + \nu^2 + \eta^2} \to \infty, \label{eq_far_field_ii}
\end{align}
for some constant $\alpha$, where $c_0(\eps \kappa)$ can be considered the electrostatic capacitance of a partially reactive disk with reactivity $\eps \kappa$. We are not aware of an explicit formula for $c_0(\eps \kappa)$. In section~\ref{kmc1}, we thus develop a numerical algorithm to calculate $c_0(\eps \kappa)$. Note that if $\eps \kappa \to \infty$, then $c_0(\eps \kappa) \to 2/\pi$, the classical electrostatic capacitance of a disk \cite{jackson-classical-1999}, matching the far-field behavior of Case I.

%%%%%%%%%%%%%%%%%%%%%%%%%%%%%%%%%%%%%%%%%%%%%%%%%%%%%%%%%%%%%%%%%%%%%%%%%%%%%%%%%
\subsubsection{Case III: $\eps\kappa\to0$}

The last case is $\eps \kappa \to 0$. We separate the analysis for this case into two subcases:
\begin{itemize}
\item \textbf{Case IIIa}: There exists some $\beta \in \N$ such that $\eps \kappa \propto \eps^\beta$.
\item \textbf{Case IIIb}: There is no $\beta \in \N$ such that $\eps \kappa \propto \eps^\beta$.
\end{itemize}

For Case IIIa, we have that there exists some constant $C>0$ such that $\eps \kappa = C \eps^\beta$. We note that this subcase includes Case 0, where neither $\kappa = o(1)$ nor $\kappa^{-1} = o(1)$ and we use an outer single perturbation expansion. We use a single perturbation in $\eps \ll 1$:
\begin{align}
w = \sum_{i=0}^\infty \eps^i w_{i}. \label{eq_expansion_iiia}
\end{align}
Substituting $C \eps^\beta$ for $\eps \kappa$, we plug this expansion into the disk boundary condition \eqref{eq_inner}:
\begin{align*}
\sum_{i=0}^\infty \eps^i \partial_\eta w_{i} = \sum_{i=0}^\infty C \eps^{i + \beta} w_{i}, \quad \eta = 0, \,\mu^2 + \nu^2 < 1.
\end{align*}
We rearrange this equation to find
\begin{align*}
\sum_{i=0}^{\beta -1} \eps^i \partial_\eta w_{i} + \sum_{i=\beta}^\infty \eps^{i} \left( \partial_\eta w_i - C w_{i-\beta} \right)= 0, \quad \eta = 0, \,\mu^2 + \nu^2 < 1.
\end{align*}
We find that the functions $w_i$ must satisfy the following boundary conditions on the disk:
\begin{align*}
\partial_\eta w_{i} &= 0, \quad i = 0, 1, \ldots, \beta-1 , \quad \eta = 0, \,\mu^2 + \nu^2 < 1,
\\
\partial_\eta w_i &= C w_{i-\beta}, \quad i = \beta, \beta+1, \ldots, \quad \eta = 0, \,\mu^2 + \nu^2 < 1.
\end{align*}
The first $\beta \geq 1$ of the $w_i$ functions experience an entirely reflective plane, so we have that these functions are constants $w_i = \alpha_i \in \R$ for $i = 0, \ldots, \beta-1$.

Next, we have
\begin{align*}
\Delta w_{\beta} &=0, \quad \eta > 0,
\\
\partial_\eta w_{\beta} &= C w_0 = C \alpha_{0}, \quad \eta = 0, \,\mu^2 + \nu^2 < 1,
\\
\partial_\eta w_{\beta} &= 0, \quad \eta = 0, \,\mu^2 + \nu^2 > 1,
\end{align*}
so $w_{\beta}$ has far-field behavior
\begin{align*}
w_{\beta} \sim - \frac{C \alpha_{0} K}{\rho}, \quad \text{ as } \rho:= \sqrt{\mu^2 + \nu^2 + \eta^2} \to \infty,
\end{align*}
for fixed constant $K$, which can be considered the electrostatic capacitance of a unit disk with fixed flux -1 (see section~\ref{kmc2} for details). We write the following general approximation for $\eps \to 0$:
\begin{align*}
w
&\sim w_{0} + \eps w_1 + \cdots + \eps^\beta w_\beta
 = \alpha + \frac{\eps \kappa}{C} w_\beta + R_a,
\end{align*}
recalling that $C \eps^\beta = \eps \kappa$ and defining $\alpha := \alpha _{0} = w_{0}$ and
\begin{align}
R_a := \sum_{i=1}^{\beta-1} \eps^i \alpha_i. \label{eq_polynomial_a}
\end{align}
Since the polynomial $R_a$ is constant in space, it will ultimately not affect the {trapping rate} (see \eqref{eq_why_polynomial} below). The far-field behavior of $w$ in this case is
\begin{align}
w \sim \alpha \left( 1 - \frac{\eps \kappa K}{\rho} \right) + R_a, \quad \text{ as } \rho:= \sqrt{\mu^2 + \nu^2 + \eta^2} \to \infty. \label{eq_far_field_iiia}
\end{align}
 
For Case IIIb, we use a double perturbation expansion in $\eps \ll 1$ and $\lambda:= \eps \kappa \ll 1$:
\begin{align}
w = \sum_{i=0}^\infty \sum_{j=0}^\infty \eps^i \lambda^j w_{i,j}. \label{eq_expansion_iiib}
\end{align}
We plug this expansion into the boundary condition on the disk \eqref{eq_inner}:
\begin{align*}
\sum_{i=0}^\infty \sum_{j=0}^\infty \eps^i \lambda^j \partial_\eta w_{i,j} = \sum_{i=0}^\infty \sum_{j=0}^\infty \eps^{i} \lambda^{j+1} w_{i,j}, \quad \eta = 0, \,\mu^2 + \nu^2 < 1.
\end{align*}
We rearrange this to find:
\begin{align*}
\sum_{i=0}^\infty \eps^i \partial_\eta w_{i,0} + \sum_{i=0}^\infty \sum_{j=1}^\infty \eps^i \lambda^j \left( \partial_\eta w_{i,j} - w_{i,j-1} \right) = 0, \quad \eta = 0, \,\mu^2 + \nu^2 < 1.
\end{align*}
Considering the hierarchy of terms, we note that the functions $w_{i,j}$ satisfy:
\begin{align*}
\partial_\eta w_{i,j} &= w_{i,j-1}, \quad i, j \geq 1, \eta = 0, \,\mu^2 + \nu^2 < 1
\\
\partial_\eta w_{i,0} &= 0, \quad \eta = 0, \,\mu^2 + \nu^2 < 1.
\end{align*}
For all $w_{i,0}$, the entire plane is reflective, so each $w_{i,0} = \alpha_i \in \R$ is constant. Next,
\begin{align*}
\Delta w_{i,1} &=0, \quad \eta > 0,
\\
\partial_\eta w_{i,1} &= \alpha_{i}, \quad \eta = 0, \,\mu^2 + \nu^2 < 1,
\\
\partial_\eta w_{i,1} &= 0, \quad \eta = 0, \,\mu^2 + \nu^2 > 1,
\end{align*}
so each $w_{i,1}$ has far-field behavior
\begin{align*}
w_{i,1} \sim - \frac{\alpha_{i} K}{\rho}, \quad \text{ as } \rho:= \sqrt{\mu^2 + \nu^2 + \eta^2} \to \infty,
\end{align*}
for the same fixed constant $K$ as in Case IIIa. Therefore, we write the following general approximation:
\begin{align*}
w &\sim w_{0,0} + \eps w_{1,0} + \lambda w_{0,1} + \eps^2 w_{2,0} + \eps \lambda w_{1,1} + \lambda^2 w_{0,2} + \cdots
\\
& = \alpha + \eps \kappa w_{0,1} + R_b,
\end{align*}
where we define $\alpha := \alpha _{0} = w_{0,0}$ and
\begin{align}
R_b := \sum_{i=1}^\infty \eps^i \alpha_i. \label{eq_polynomial_b}
\end{align}
Since the polynomial $R_b$ is constant in space, it will ultimately not affect the {trapping rate} (see \eqref{eq_why_polynomial} below). The far-field behavior of $w$ in this case is
\begin{align}
w \sim \alpha \left( 1 - \frac{\eps \kappa K}{\rho} \right) + R_b, \quad \text{ as } \rho:= \sqrt{\mu^2 + \nu^2 + \eta^2} \to \infty, \label{eq_far_field_iiib}
\end{align}
which matches the far-field behavior of $w$ in \eqref{eq_far_field_iiia} for Case IIIa, using polynomial $R_b$ in \eqref{eq_polynomial_b} instead of polynomial $R_a$ in \eqref{eq_polynomial_a}.

\subsection{Matching}

The leading order far-field behavior for each case is
\begin{align}
w \sim \alpha \left( 1 - \frac{ G_{i}(\eps \kappa) }{\rho} \right) + R_{i}, \text{ as } \rho := \sqrt{\mu^2 + \nu^2 + \eta^2} \to \infty, \label{eq_inner_farfield}
\end{align}
where $\alpha$ is a constant determined by matching with the outer solution, $G_{i}(\eps \kappa)$ is a function of $\eps \kappa$ depending on which case $i\in\{\text{I},\text{II},\text{III}\}$ applies, and $R_{i}$ is a polynomial in $\eps$ for Case III and $R_{i}=0$ for Case I and Case II. The forms of $G_{i} (\eps \kappa)$ are:
\begin{align*}
G_\text{I}(\eps \kappa) &= 2/\pi,
\\
G_\text{II}(\eps \kappa) &= c_0(\eps \kappa),
\\
G_\text{III}(\eps \kappa) &= \eps \kappa K.
\end{align*}
We will show that $\alpha$ is the same for each case.

Our matching condition is that the near-field behavior of the outer expansion as $\mathbf{x} \to ({x_{n}}, {y_{n}}, 0)$ must agree with the far-field behavior of the inner expansion as $\rho \to \infty$. That is, for the case of the outer double perturbation expansion,
\begin{align}
1 +& \eps S_{1,0} + \kappa^{\gamma} S_{0,1} + \eps^2 S_{2,0} + \eps \kappa^{\gamma} S_{1,1} + \kappa^{2\gamma} S_{0,2} + \cdots \nonumber
\\
& \sim \alpha \left( 1 - \frac{G_{i}(\eps \kappa)}{\rho} \right) + R_{i},\quad \mathbf{x}\to ({x_{n}}, {y_{n}}, 0), \rho \to \infty. \label{eq_s_outer_expansion}
\end{align}
For all cases, $\alpha = 1$, and $\gamma \in \{-1,1\}$ depends on the size of $\kappa$, as used in the outer expansion \eqref{eq_outer_expansion}. For Cases I and II, the second term has order $\eps$, so we determine the singular behavior as $\mathbf{x} \to \mathbf{x}_n$ of $S_{1,0}$ to be
\begin{align*}
S_{1,0} \sim \frac{- G_{i}(\eps \kappa)}{\left| \mathbf{x} - \mathbf{x}_n \right| }.%,
\end{align*}

For Case III, the first term that is not constant in space in the inner expansion has order $\eps^2 \kappa$, and the outer expansion uses $\gamma = 1$, so we determine that the second term in the outer expansion is $\eps^2 \kappa S_{2,1}$ which has singular behavior
\begin{align*}
S_{2,1} \sim \frac{-K}{\left| \mathbf{x} - \mathbf{x}_n \right| }\quad\text{as $\mathbf{x} \to \mathbf{x}_n$}.
\end{align*}
Aside from the polynomials $R_{i}$ in Case III given by \eqref{eq_polynomial_a} and \eqref{eq_polynomial_b} that are constant in space and do not affect the {trapping rate}, we can ignore all smaller terms.

Similarly, for Case 0 with the outer single perturbation expansion, our matching condition is also that the near-field behavior of the outer expansion as $\mathbf{x} \to ({x_{n}}, {y_{n}}, 0)$ must agree with the far-field behavior of the inner expansion as $\rho \to \infty$. That is,
\begin{align}
1 +& \eps S_{1} + \eps^2 S_{2} + \cdots \nonumber
\\
& \sim \alpha \left( 1 - \frac{\eps \kappa K}{\rho} \right) + R_{i},\quad \mathbf{x}\to ({x_{n}}, {y_{n}}, 0), \rho \to \infty. \label{eq_s_outer_expansion_single}
\end{align}
We again find $\alpha=1$. The first term that is not constant in space in the inner expansion has order $\eps^2$, so we determine that the second term in the outer expansion is $\eps^2 S_2$, which has singular behavior
\begin{align*}
S_2 \sim \frac{-\kappa K}{\left| \mathbf{x} - \mathbf{x}_n \right| }\quad\text{as $\mathbf{x} \to \mathbf{x}_n$}.
\end{align*}
Again aside from the polynomials $R_{i}$ in Case III given by \eqref{eq_polynomial_a} and \eqref{eq_polynomial_b} that are constant in space and do not affect the {trapping rate}, we can ignore all smaller terms.

For all cases, we can write the singular behavior of the second term, denoted by $S_* \in \{S_{1,0}, S_{2,1}, S_2 \}$, for each $n \in \N$ in the distributional form to derive the following general boundary value problem for $S_{1,0}$, $S_{2,1}$, and $S_2$:
\begin{align}\begin{split}
\partial_t S_* &= \Delta S_*, \quad z>0,\, t > 0,
\\
\partial_z S_* &= 2 \pi G_{i}(\eps \kappa) \sum_{n=1}^\infty \delta(x-{x_{n}}) \delta(y-{y_{n}}), \quad z = 0.\label{distform}
\end{split}
\end{align}

To derive the distributional form in \eqref{distform}, suppose a function $f$ satisfies \eqref{distform} and define the inner solution
\begin{align*}
h(\mu, \nu, \eta, t) := f \left( \eps \mu + {x_{n}}, \eps \nu + {y_{n}}, \eps \eta, t \right).
\end{align*}
We introduce the inner expansion $h \sim h_0 + \cdots$ and find that $h_0$ is harmonic in $(\mu, \nu, \eta)$,
\begin{align*}
\Delta h_0 = 0, \quad (\mu, \nu) \in \R^2, \eta > 0.
\end{align*}
Then we derive the boundary condition for $h_0$ at $\eta = 0$:
\begin{align*}
\eps^{-1} \partial_\eta h = \partial_z f &=2 \pi G_{i}(\eps \kappa) \sum_{n=1}^\infty \delta(x-{x_{n}}) \delta(y-{y_{n}})
\\
&= 2 \pi \eps^{-2} G_{i}(\eps \kappa) \delta(\mu) \delta(\nu),
\end{align*}
so we have
\begin{align*}
\partial_\eta h_0 = 2 \pi \eps^{-1} G_{i}(\eps \kappa) \delta(\mu) \delta(\nu).
\end{align*}
We use the half-space three-dimensional Green's function $\mathcal{G}(a,b) = -(2 \pi | a-b|)^{-1}$ satisfying
\begin{gather*}
\Delta \mathcal{G}(a,b) =\delta(a-b),
\end{gather*}
to conclude that the exact solution for $h_0$ is
\begin{align*}
h_0 = \Lambda - \frac{\eps^{-1} G_{i}(\eps \kappa)}{\sqrt{ \mu^2 + \nu^2 + \eta^2}}.
\end{align*}
This solution matches the far-field behavior of $w$ in \eqref{eq_inner_farfield}. For $\Lambda = 0$, we have
\begin{align*}
h \sim h_0 = - \frac{\eps^{-1} G_{i}(\eps \kappa)}{\sqrt{ \mu^2 + \nu^2 + \eta^2}} = \frac{ G_{i}(\eps \kappa)}{\left| \mathbf{x} - \mathbf{x}_n \right|},
\end{align*}
confirming the singular behavior of $S_*$.

%%%%%%%%%%%%%%%%%%%%%%%%%%%%%%%%%%%%%%%%%%%%%%%%%%%%%%%%%%%%%%%%%%%%%%%%%%%%%%%%
\subsection{Homogenization} \label{sec_homogenization}

%%%%%%%%%%%%%%%%%%%%%%%%%%%%%%%%%%%%%%%%%%%%%%%%%%%%%%%%%%%%
We define the homogenized survival probability $\overline{S}(z,t)$ by conditioning that $Z(0)=z\ge0$ and averaging over the positions of $x$ and $y$ in the plane,
\begin{align}\label{eq_sbar_definition}
\overline{S}(z,t)
:=\lim_{l\to\infty}\frac{1}{(2l)^{2}}\int_{-l}^{l}\int_{-l}^{l}S(x,y,z,t)\,\dd x\,\dd y.
\end{align}
It follows from \eqref{eq_s_all} that $\overline{S}$ satisfies the boundary value problem
\begin{align*}
\partial_t \overline{S} &= \partial_{zz} \overline{S}, \quad t>0,
\\
\lim_{z \to \infty} \overline{S}(z,t) &= 1, \quad t > 0,
\\
\overline{S}(z,t) &= 1, \quad t=0,\,z > 0.
\end{align*}
We want to derive the Robin boundary condition that $\overline{S}$ satisfies at $z=0$, so we define the ratio
\begin{align}
\chi = \frac{\partial_z \overline{S} (0, t)}{\overline{S}(0,t)} > 0, \label{eq_chi_definition}
\end{align}
which we will show is independent of time to leading order as $\eps\to0$. Given the definition of $\chi$ in \eqref{eq_chi_definition}, it is a tautology that $\overline{S}$ satisfies the following boundary condition at $z=0$,
\begin{align*}
\partial_z \overline{S} = \chi \overline{S}, \quad z = 0.
\end{align*}

The denominator in the definition of $\chi$ in \eqref{eq_chi_definition} approaches unity as $\eps\to0$ by \eqref{obv1}. To determine the behavior of the numerator in \eqref{eq_chi_definition}, we  interchange the derivative and integrals, substitute in the outer expansion for $S$ given in \eqref{eq_s_outer_expansion}, and take the leading order term as $\eps \to 0$, where we let $m=0$ for Cases I and II and $m=1$ for Case III:
\begin{subequations}\label{eq_why_polynomial}
\begin{align}
\chi
&\sim \eps (\eps \kappa)^m \lim_{l\to\infty}\frac{1}{(2l)^{2}}\int_{-l}^{l}\int_{-l}^{l} \partial_z S_*(\mathbf{x},t) \, \dd x\, \dd y \Big|_{z=0}\quad\text{as }\eps\to0. %\\
%& \qquad \sim \eps (\eps \kappa)^m \int_{-\frac{1}{2}}^{\frac{1}{2}}\int_{-\frac{1}{2}}^{\frac{1}{2}} \partial_z S_*(\mathbf{x},t) \, dx dy \Big|_{z=0}, \quad \eps \to 0.
\end{align}
\end{subequations}
Note that the leading order term in the numerator involves $\partial_z S_*(\mathbf{x},t)$ for all cases, including Case III, where $\partial_z R_{i} = 0$ for the polynomials $R_{i}$ in $\eps$ in \eqref{eq_polynomial_a} and \eqref{eq_polynomial_b}. Using the boundary condition in \eqref{distform} then implies
\begin{align}
\chi & \sim \eps (\eps \kappa)^m 2 \pi G_{i}(\eps \kappa) \lim_{l\to\infty}\frac{1}{(2l)^{2}}\int_{-l}^{l}\int_{-l}^{l}   \sum_{n=1}^\infty \delta(x-{x_{n}}) \delta(y-{y_{n}}) \, \dd x\, \dd y, \quad \eps \to 0, \nonumber
\\
& =2 \pi \eps (\eps \kappa)^m G_{i}(\eps \kappa), %\quad \eps \to 0,
\label{eq_chi_c0}
\end{align}
where we have used the assumption in \eqref{mr} on the absorbing surface fraction and the rescaling in \eqref{rescaling}.
To summarize, the trapping rates for the three cases are:
\begin{align}
\chi = \begin{cases}
4 \eps, \quad & \text{Case I: } \eps \kappa \gg 1,
\\
2 \pi \eps c_0(\eps \kappa), & \text{Case II: } \eps \kappa = O(1),
\\
2 \pi \eps^2 \kappa K, & \text{Case III: } \eps \kappa \ll 1.
\end{cases} \label{eq_reacs}
\end{align}
Note that the trapping rate for Case I of $\chi=4\eps$ is the trapping rate for perfectly absorbing patches (i.e.\ $\kappa=\infty$). Table \ref{table_sum} summarizes the above results and \eqref{main} gives the trapping rate in dimensional units (i.e.\ $\overline{\kappa}=\chi D/L$).

%%%%%%%%%%%%%%%%%%%%%%%%%%%%%%%%%%%%%%%%%%%%%%%%%%%%%%%%%%%%%%%%%%%%%%%%%%%%%%%%%%%%%%%%%%%%%%%%%%%%%%%%%%%%%%%%%%%%%%%%%%%%%%%%%%%%%%%%%%%%%%%%%%%%%%%%%%%%%%%%%%%%%%%%%%%%%%%%%%%%%%%%%%%%%%%%%%%%%%%%%%%%%%%%%%%%%%%%%%%%%%%%%
\section{KMC for a single partially reactive patch}\label{kmc1}

In this section, we develop a KMC algorithm to calculate the capacitance $c_0(\eps \kappa)$ of a single partially reactive disk. In section~\ref{kmc2}, we develop a similar KMC algorithm to calculate the capacitance $K$ of disk with a fixed flux. In section~\ref{kmc3}, we develop a similar KMC algorithm to confirm the results of the analysis in section~\ref{math} by simulating the full stochastic system.

These algorithms break the simulation process of a Brownian particle diffusing above a partially reactive surface into two simple diffusion stages (from bulk to boundary and from boundary to bulk) and alternate between these two stages until reaching a breaking point. These simple diffusion stages are on simple subdomains (hemispheres, disks, semi-infinite intervals) and can thus be exactly simulated because the corresponding PDEs on these simple subdomains can be solved analytically. These stages then involve large step sizes, allowing for greater computational efficiency compared to traditional Brownian motion simulation methods. By simulating many $M \gg 1$ realizations of these Brownian particles, we can produce accurate approximations of constants, including capacitances and effective trapping rates. These algorithms modify a method devised by Bernoff, Lindsay, and Schmidt \cite{bernoff-boundary-2018} for perfectly absorbing patches. The basic idea of the algorithms date back to the so-called walk-on-spheres method of Muller \cite{muller1956}.

%%%%%%%%%%%%%%%%%%%%%%%%%%%%%%%%%%%%%%%%%%%%%%%%%%%%%%%%%%%%%%%%%%%%%%%%%%%%%%%%%%%%%%%%%%%%%%%%%%%%%%%%%
\subsection{Probabilistic representation}

%%%%%%%%%%%%%%%%%%%%%%%%%%%%%%%%%%%%%%%%%%%%%%%%%%%%%%%%%%%%%%%%%%%%%%%%%%%%%%%%%%%%%%%%%%%%%%%%%%%%%%%%%%%%%%%%%%%%%%%%%%%%%%%%%%%%%%%%%%%%%%%%

The far-field behavior in \eqref{eq_far_field_ii} of the inner expansion given in section~\ref{sec_inner} for Case II relies on the electrostatic capacitance $c_0(\eps \kappa)$ of a single partially reactive disk with reactivity $\eps \kappa$ and unit radius. For clarity, we define ${{\kappa'}} := \eps \kappa$. We use a probabilistic representation of the leading order inner solution $w_{0,0}$ from section~\ref{sec_inner} Case II to calculate the electrostatic capacitance $c_0({{\kappa'}})$. 

Let $\mathbf{X}(t) \in \R^3$ be a standard three-dimensional Brownian particle with unit diffusion coefficient and define $\tau$ to be the time of absorption in the partially reactive disk with reactivity ${{\kappa'}}$ on the otherwise reflective plane as described by the boundary conditions \eqref{eq_inner_bc1} and \eqref{eq_inner_bc2}. The leading order inner solution $w_{0,0}$ that is harmonic in upper half-space and satisfies the boundary conditions \eqref{eq_inner_bc1} and \eqref{eq_inner_bc2} can be written as
\begin{align*}
w_{0,0}(\mu, \nu, \eta) = 1- q(\mu, \nu, \eta),
\end{align*}
where $q$ is the probability that the Brownian particle $\mathbf{X}$ is eventually absorbed by the disk, conditioned on the initial position $\mathbf{X}(0)$:
\begin{align*}
q(\mu, \nu, \eta) = \P ( \tau < \infty \, | \, \mathbf{X}(0) = (\mu, \nu, \eta) ).
\end{align*}
Since $w_{0,0}$ is harmonic for $\eta > 0$, $q$ is also harmonic for $\eta > 0$ and has the boundary conditions
\begin{align*}
\partial_\eta q = {{\kappa'}} (q-1), &\quad \eta = 0, \,\mu^2 + \nu^2 < 1,
\\
\partial_\eta q = 0, &\quad \eta = 0, \,\mu^2 + \nu^2 > 1.
\end{align*}
Define $\overline{q}(\rho)$ to be the average of $q$ over the surface of the hemisphere of radius $\rho$ centered at the origin. If $\rho > 1$, then $\partial_\eta q = 0$ for $\eta = 0$. We can therefore integrate the harmonic PDE for $q$ over the surface of the hemisphere, and by using the divergence theorem and interchanging integration and differentiation, we have that $\overline{q}(\rho)$ satisfies
\begin{align}\label{qnice}
\Big(\frac{2}{\rho} \partial_\rho + \partial_\rho^2 \Big) \overline{q} = 0, \quad \rho > 1.
\end{align}
Solving this and matching with the far-field behavior of $w_{0,0}$ in \eqref{eq_inner_farfield}, we have
\begin{align*}
\overline{q}(\rho) = \frac{c_0({{\kappa'}})}{\rho}, \quad \rho > 1.
\end{align*}
Therefore, if we know the probability $\overline{q}(\rho)$ that a Brownian particle $\mathbf{X}$ starting uniformly on the surface of the hemisphere of radius $\rho > 1$ is absorbed by the partially reactive disk of unit radius with reactivity ${{\kappa'}}$, then we can calculate the capacitance $c_0({{\kappa'}})$. We estimate the probability $\overline{q}(\rho)$ by simulating $M \gg 1$ realizations of a Brownian particle $\mathbf{X}$ and calculating the proportion of realizations absorbed before reaching a large outer radius $\rho_\infty \gg 1$.

%%%%%%%%%%%%%%%%%%%%%%%%%%%%%%%%%%%%%%%%%%%%%%%%%%%%%%%%%%%%%%%%%%%%%%%%%%%%%%%%%%%%%%%%%%%%%%%%%%%%%%%%%
\subsection{Simulation algorithm}

We simulate these realizations via the following {{KMC}} algorithm. The idea of the algorithm is to break up the simulation into smaller steps involving simple domains wherein the motion of the particle can be simulated exactly because the corresponding PDEs can be solved on these simple domains.

Specifically, each simulation begins by placing a Brownian particle $\mathbf{X}$ on the hemisphere centered at the origin of radius $\rho > 1$ in upper half-space, according to a uniform distribution on the surface. The simulation proceeds by alternating between the following stages, extending the two stages developed by Bernoff, Lindsay, and Schmidt \cite{bernoff-boundary-2018} for the problem of a perfectly reactive patch to account for the partially reactive patch. In particular, the algorithm is identical to that in \cite{bernoff-boundary-2018} except for Stage IIA (the algorithm in \cite{bernoff-boundary-2018} would terminate at our Stage IIA).

\begin{itemize}
\item \textit{Stage I: Project from bulk to plane:} The particle is projected to the $\eta = 0$ plane using the exact distribution. If the particle lands on the plane such that it is within the disk, the algorithm proceeds to Stage IIA. If the particle lands on the plane outside of the disk, then the algorithm proceeds to Stage IIB.
\item \textit{Stage II: Project from plane to bulk:} Depending on where on the plane the particle landed, the particle either experiences a partially reactive or reflecting boundary.
\begin{itemize}
\item \textit{Stage IIA: Within partially reactive patch:} In the case that the particle lands on the plane within the partially reactive patch, the algorithm computes the distance $d_1$ to the boundary of the patch and then samples the time $t$ until $(\mu, \nu)$ diffuses a distance $d_1$. The algorithm samples if the patch absorbs the particle before time $t$ has elapsed. If the particle is absorbed, we record a success, and the trial ends. Otherwise, the algorithm samples the particle's location in the bulk and returns to Stage I.
\item \textit{Stage IIB: Outside partially reactive patch:} The algorithm computes the distance $d_2$ to the patch boundary. The particle is projected onto the hemisphere of radius $d_2$ following a uniform distribution. If the particle is farther than $\rho_\infty$ from the origin, we record a failure, and the trial ends. Otherwise, the algorithm returns to Stage I.
\end{itemize}
\end{itemize}

The distribution of the particle on the plane at the end of Stage I is found by sampling the random time that it takes the particle to reach the $\eta = 0$ plane from the current location $\mathbf{X} = (\mu, \nu, \eta)$, by calculating \cite{bernoff-boundary-2018}
\begin{align*}
t^* = \frac{1}{4} \left( \frac{\eta}{\text{erfc}^{-1}(U)}\right)^2,
\end{align*}
where $U$ is a uniform random variable on $[0,1]$. Then the location at the end of Stage I is given by
\begin{align*}
(\mu, \nu, 0) + \sqrt{2 t^*} (\xi_1, \xi_2, 0),
\end{align*}
where $\xi_1$ and $\xi_2$ are independent standard normal random variables.

For Stage IIA, we wish to determine if the particle is absorbed by the partially reactive boundary while diffusing above it. Following the particle's first encounter with the partially reactive boundary, it experiences many more encounters before diffusing away. Since the particle diffuses isotropically, we can separately consider diffusion in the $\eta$ direction and the $(\mu, \nu)$ directions, allowing for greater computational efficiency. The algorithm first samples the time $t^*$ that it takes to diffuse to a circle of radius $d_1$ in the $(\mu, \nu)$ plane, where $d_1$ is the distance to the boundary. Ciesielski and Taylor \cite{ciesielski-first-1962} give a long-time expansion for the cumulative distribution function of the first exit time from the unit disk as
\begin{align}
P_c(s) = \sum_{r=1}^\infty \frac{ \exp \left( - q_{r}^2 s /2 \right)}{q_{r} J_1(q_{r})}, \label{cyl_cdf}
\end{align}
where $J_i(z)$ are the $i$th-order Bessel functions of the first kind and $q_r$ are the ordered positive roots of $J_0(z)$. We precompute the cumulative distribution function $P_c(s)$ using the first $2 \times 10^7$ terms in the series \eqref{cyl_cdf}, using the \texttt{besselzero} function \cite{nicholson-bessel-2018} in MATLAB \cite{matlab-91001739362-2021}. The algorithm samples a uniform random number $\xi \in [0,1]$, numerically solves for $s^*$ satisfying
\begin{align*}
P_c(s^*) = \xi,
\end{align*}
and then calculates the exit time $t^* = d_1^2 s^*$.

From the first encounter with the plane at the start of Stage IIA until $t^*$ time has elapsed, the particle is guaranteed to experience the partially reactive boundary in the $\eta$ direction. We wish to determine where the particle is in the $\eta$ direction at time $t^*$ or if it was absorbed before time $t^*$. To do so, we sample from the mixed state space that contains a discrete state $\mathghost$ denoting absorption and a continuous space $[0, \infty)$ denoting location if not yet absorbed:
\begin{align*}
\mathcal{S} = \{ \mathghost \} \cup [0, \infty). %\label{eq_state_space}
\end{align*}
Using the solution to the 1D diffusion on a line with a partially absorbing boundary at $\eta = 0$ given by Carslaw and Jaeger \cite{carslaw-conduction-1959}, we calculate the partial cumulative distribution for the location $\eta \in [0, \infty)$ after time $t^*$:
\begin{align}
P_p(\eta, t^*, {{\kappa'}}) = e^{ {{\kappa'}}^2 t^*} \text{erfc}\left( {{\kappa'}} \sqrt{t^*}\right) - e^{{{\kappa'}} ({{\kappa'}} t^* + \eta) }\text{erfc} \left( \frac{2 {{\kappa'}} t^* + \eta}{2 \sqrt{t^*}} \right). \label{eq_absorb_loc}
\end{align}
In words, $P_{p}(\eta,t^{*},{{\kappa'}})$ is the probability that at time $t^{*}$, the particle has not absorbed and is located at a height below $\eta$. 
Hence, taking $\eta\to\infty$ gives the probability that the particle is absorbed before time $t^{*}$. Denoting this absorption probability by $Q(t^{*},{{\kappa'}})$, using \eqref{eq_absorb_loc} yields the following formula
\begin{align}
Q(t^*, {{\kappa'}})
=1-\lim_{\eta \to \infty} P_p(\eta, t^*, {{\kappa'}})
= \exp \left({{\kappa'}}^2 t^* \right) \text{erfc} \left( {{\kappa'}} \sqrt{t^*} \right). \label{eq_p_ghost}
\end{align}

The algorithm samples a uniform random number $\xi \in [0,1]$ and attempts to numerically solve for $\eta$ using
\begin{align*}
P_p(\eta,t^*, {{\kappa'}}) = \xi.
\end{align*}
If the random number $\xi$ is greater than or equal to the probability of not being absorbed, (meaning, $\xi \geq 1 - Q(t^*, {{\kappa'}})$ using \eqref{eq_p_ghost}), then this does not return a value for $\eta$. A value not being returned indicates that the particle is absorbed. If the particle is absorbed, the algorithm records a success. Otherwise, the algorithm updates the position of the particle as
\begin{align*}
(\mu, \nu, 0) + (d_1 \cos \theta, d_1 \sin \theta, \eta),
\end{align*}
where $\theta$ is sampled uniformly on $[ 0, 2 \pi]$ and $\eta$ was sampled above.

For Stage IIB, the distance $d_2$ to the boundary of the patch is calculated. The furthest that the particle can diffuse without encountering the partially reactive boundary is $d_2$, so the algorithm propagates the particle to the hemisphere of radius $d_2$ according to
\begin{align*}
(\mu, \nu, 0) + \frac{d_2}{\sqrt{\xi_1^2 + \xi_2^2 + \xi_3^2}}(\xi_1, \xi_2, |\xi_3|).
\end{align*}
The algorithm calculates the distance from the origin, and if it is greater than $\rho_\infty$, the algorithm records a failure, and the trial ends. Otherwise, the algorithm returns to Stage I.

%%%%%%%%%%%%%%%%%%%%%%%%%%%%%%%%%%%%%%%%%%%%%%%%%%%%%%%%%%%%%%%%%%%%%%%%%%%%%%%%%%%%%%%%%%%%%%%%%%%%%%%%%
\subsection{Simulation results}\label{sigmoid}

%%%%%%%%%%%%%%%%%%%%%%%%%%%%%%%%%
\begin{figure}%[t]
\centering
(a)\hspace{-.05\textwidth}\includegraphics[width=.47\textwidth]{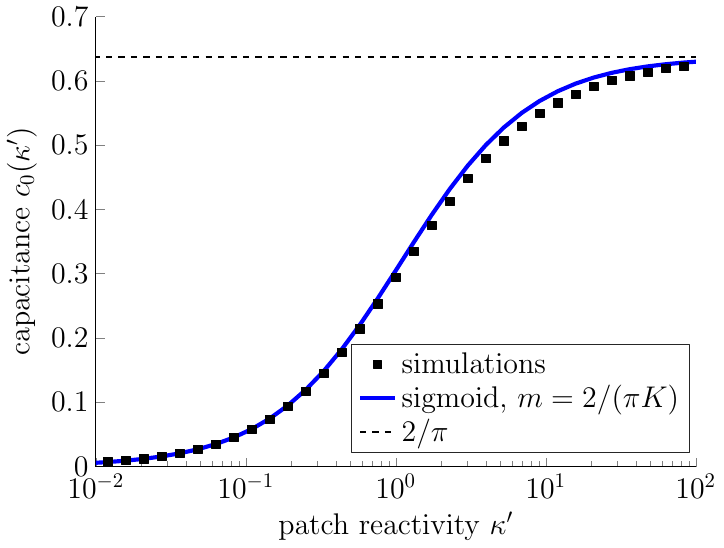}
\hspace{.05\textwidth}
(b)\hspace{-.05\textwidth}\includegraphics[width=.47\textwidth]{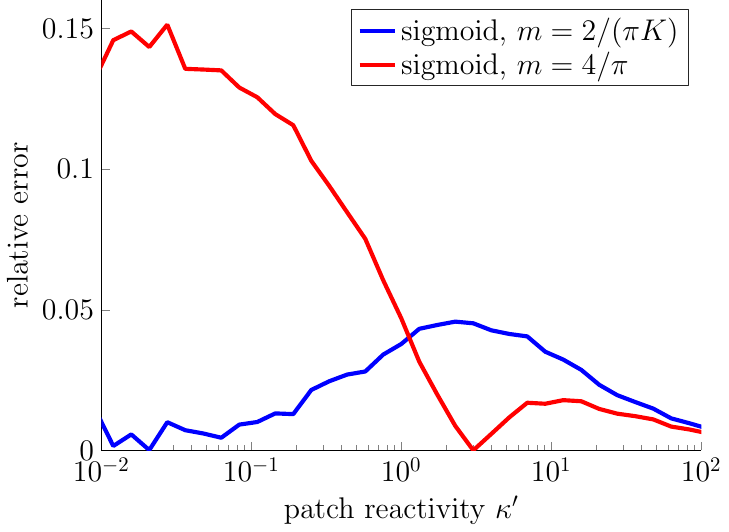}
\caption[Numerically-calculated capacitance of a unit disk for varying reactivity ${{\kappa'}}$.]
{Numerically-calculated capacitance $c_{0}(\kappa')$ of a unit disk for varying reactivity ${{\kappa'}}$. (a): Capacitance as a function of ${{\kappa'}} \in [10^{-2}, 10^{2}]$, for $10^6$ trials with a maximum radius of $\rho_\infty = 10^{10}$, and the sigmoidal estimate $c_{\textup{sig}}(\kappa',m)$ in \eqref{eq_c0_approx} with $m=2/(\pi K)$ in \eqref{mest} with $K \approx 0.{5854}$. (b): Relative error, $| c_0({{\kappa'}}) - c_{\textup{sig}}(\kappa',m) |/c_0({{\kappa'}})$, of the sigmoidal approximation for $m=2/(\pi K)$ and $m=4/\pi$. }
\label{figc0}
\end{figure}
%%%%%%%%%%%%%%%%%%%%%%%%%%%%%%%%%

In Figure~\ref{figc0}a, we plot the capacitance $c_{0}(\kappa')$ computed from  this KMC simulation algorithm (black square markers). For each value of ${{\kappa'}}$ simulated, $10^6$ trials were performed with a maximum radius of $\rho_\infty = 10^{10}$ and the proportion $\overline{q}(\rho)$ of trials which ended with absorption in the disk was recorded. The capacitance was then computed via $c_0({{\kappa'}}) = \rho \overline{q}(\rho)$. Note that as ${{\kappa'}} \to \infty$, the capacitance $c_0({{\kappa'}}) \to 2/\pi$ as expected \cite{bernoff-boundary-2018}. The results in Figure~\ref{figc0}a suggest that a sigmoid function is a possible approximation of $c_0({{\kappa'}})$:
\begin{align}
c_0({{\kappa'}}) \approx c_{\textup{sig}}(\kappa',m)
:=\frac{(2/\pi) {{\kappa'}}}{{{\kappa'}} + m}, \label{eq_c0_approx}
\end{align}
where $m$ is an appropriately chosen constant. The blue curve in Figure~\ref{figc0}a is the sigmoid approximation in \eqref{eq_c0_approx} with
\begin{align}\label{mest}
m=2/(\pi K),
\end{align}
where $K \approx 0.{5854}$ is calculated in Section~\ref{kmc2}. The choice \eqref{mest} stems from noting that (i) $c_{\textup{sig}}(\kappa',m)\sim 2/(\pi m)\kappa'$ as $\kappa'\to0$ and (ii) we must have $c_0({{\kappa'}})\sim \kappa' K$ as $\kappa'\to0$ if Case II is to agree with Case III in \eqref{eq_reacs} for $\eps\kappa\ll1$.

The relative error between the $c_{0}(\kappa')$ and the sigmoid approximation $ c_{\textup{sig}}(\kappa',m)$ with $m=2/(\pi K)$ is shown in Figure~\ref{figc0}b (blue curve). This relative error reaches a maximum of less than $5\%$ for $\kappa'\approx2$. Figure~\ref{figc0}b also shows the relative error for the sigmoid approximation with $m=4/\pi$, where choosing $m=4/\pi$ stems from a prior heuristic trapping rate formula (see Section~\ref{comparison} for details).

%%%%%%%%%%%%%%%%%%%%%%%%%%%%%%%%%%%%%%%%%%%%%%%%%%%%%%%%%%%%%%%%%%%%%%%%%%%%%%%%%%%%%%%%%%%%%%%%%%%%%%%%%%%%%%%%%%%%%%%%%%%%%%%%%%%%%%%%%%%%%%%%
\section{KMC for a single patch with fixed flux} \label{kmc2}

We now develop a KMC algorithm to compute the capacitance of a single patch with a fixed flux. As in Section~\ref{kmc1}, our algorithm relies on a probabilistic representation of the solution to a certain PDE. In this case, the probabilistic representation involves Brownian local time.

%%%%%%%%%%%%%%%%%%%%%%%%%%%%%%%%%%%%%%%%%%%%%%%%%%%%%%%%%%%%%%%%%%%%%%%%%%%%%%%
\subsection{Probabilistic representation using local time} \label{app_local}

The far-field behavior in \eqref{eq_far_field_iiia} and \eqref{eq_far_field_iiib} for Case III relies on the capacitance $K$ of a unit disk with a fixed-flux boundary condition. This arises in the far-field behavior of the PDE:
\begin{align}\label{fpde}
\begin{split}
\Delta f= 0, &\quad \eta > 0, (\mu, \nu) \in \R^2,
\\
\partial_\eta f = 1, &\quad \eta = 0, \,\mu^2 + \nu^2 < 1,
\\
\partial_\eta f = 0, &\quad \eta = 0, \,\mu^2 + \nu^2 > 1,
\\
\lim_{\rho \to \infty} f = 0, &\quad \rho = \sqrt{\mu^2 + \nu^2 + \eta^2}.
\end{split}
\end{align}
Let $\overline{f}(\rho)$ be the average of $f(\mu, \nu, \eta)$ over the surface of the hemisphere of radius $\rho := \sqrt{\mu^2 + \nu^2 + \eta^2} > 0$ with $\eta \geq 0$. As in \eqref{qnice}, integrating $\Delta f$ over the hemisphere of radius $\rho > 1$ and using the divergence theorem yields
\begin{align*}
\Big(\frac{2}{\rho} \partial_\rho + \partial_\rho^2 \Big) \overline{f} = 0, \quad \rho > 1.
\end{align*}
Given the outer limit $\lim_{\rho \to \infty} f(\mu, \nu, \eta) = 0$, we have that the general solution is $\overline{f}(\rho) = -K \rho^{-1}$ for some constant $K \in \R$.

To calculate $K$, we first claim that $f$ satisfies
\begin{align}\label{claim78}
f(\mu, \nu, \eta) = -\E [ \ell_\infty \, | \, \mathbf{X}(0) = (\mu, \nu, \eta) ],
\end{align}
where $\ell_\infty = \lim_{t\to \infty} \ell(t)$ is the limit of the boundary local time $\ell(t)$ on the unit disk centered at the origin accumulated by a Brownian particle $\{\mathbf{X}(t)\}_{t\ge0}$ in upper half-space with reflection at the $\eta=0$ plane. To obtain \eqref{claim78}, suppose $f$ satisfies \eqref{fpde} and apply Ito's formula \cite{bass1998} to yield 
\begin{align}\label{ito}
\begin{split}
f(\mathbf{X}(t\wedge\tau_{R}))
&=f(\mathbf{X}(0))
+M
+\frac{1}{2}\int_{0}^{t\wedge\tau_{R}}\Delta f(\mathbf{X}(s))\,\dd s\\
&\quad+\int_{0}^{t\wedge\tau_{R}}\partial_{\eta}f(\mathbf{X}(s))\,\dd \ell(s)
+\int_{0}^{t\wedge\tau_{R}}\partial_{\eta}f(\mathbf{X}(s))\,\dd \ell_{0}(s),
\end{split}
\end{align}
where $M$ is a martingale, $t\wedge\tau_{R}$ denotes the minimum of $t>0$ and $\tau_{R}>0$, where $\tau_{R}$ is the first time the Brownian particle escapes the sphere of radius $R>0$,
\begin{align*}
\tau_{R}
:=\inf\{t>0:\|\mathbf{X}(t)\|>R\},
\end{align*}
and $\ell_{0}(t)$ is the boundary local time at the $\eta=0$ plane away from the unit disk centered at the origin (i.e.\ $\ell(t)+\ell_{0}(t)$ is the total local time at the $\eta=0$ plane). Using that $f$ satisfies \eqref{fpde} and taking the expectation of \eqref{ito} conditioned that $\mathbf{X}(0)=(\mu,\nu,\eta)$ and taking $t\to\infty$ yields
\begin{align}\label{lhsito}
\E[f(\mathbf{X}(\tau_{R}))\,|\,\mathbf{X}(0)=(\mu,\nu,\eta)]
=f(\mu,\nu,\eta)+\E[\ell(\tau_{R})\,|\,\mathbf{X}(0)=(\mu,\nu,\eta)].
\end{align}
The lefthand side of \eqref{lhsito} vanishes as $R\to\infty$ by the far-field condition in \eqref{fpde}. Further, $\tau_{R}\to\infty$ almost surely as $R\to\infty$, and thus \eqref{lhsito} implies \eqref{claim78}.

\subsection{Simulation algorithm for local time}

We estimate the value of $\overline{f}(\rho)$ by simulating $M \gg 1$ realizations of a Brownian particle $\mathbf{X}$ and calculating the average of the realizations of accumulated local time before the particle reaches a large outer radius $\rho_\infty \gg 1$. The {{KMC}} algorithm to sample the local time consists of two stages, where one stage has two cases. Stage I and Stage IIB are nearly identical to the algorithm for a partially reactive patch in section~\ref{kmc1}, and Stage IIA is the stage where the particle accumulates local time.

As in the partially reactive patch algorithm, each simulation begins by placing a Brownian particle $\mathbf{X}$ on the hemisphere of radius $\rho > 1$ in upper half-space, according to a uniform distribution on the surface. The total local time $L$ is initialized at 0. The simulation proceeds by alternating between the following stages, allowing local time to accumulate until a breaking point.
\begin{itemize}
\item \textit{Stage I: Project from bulk to plane:} The particle is projected to the $\eta = 0$ plane using the exact distribution. If the particle lands on the plane such that it is within the disk, the algorithm proceeds to Stage IIA. If the particle lands on the plane outside of the disk, then the algorithm proceeds to Stage IIB.
\item \textit{Stage II: Project from plane to bulk:} Depending on where on the plane the particle landed, the particle either accumulates local time or does not before being reflected.
\begin{itemize}
\item \textit{Stage IIA: Within Patch:} In the case that the particle lands on the plane within the patch, the algorithm computes the distance $d_1$ to the boundary of the patch and then samples the time $t$ until $(\mu, \nu)$ diffuses a distance $d_1$. The accumulated local time $\ell$ within the time $t$ is sampled and added to the total local time $L$. The particle's location in the bulk is sampled based on $t$ and $\ell$, and the algorithm returns to Stage I.
\item \textit{Stage IIB: Outside Patch:} The algorithm computes the distance $d_2$ to the boundary of the patch. The particle is projected onto the hemisphere of radius $d_2$ following a uniform distribution. If the particle is farther than $\rho_\infty$ from the origin, we record the total local time $L$, and the trial ends. Otherwise, the algorithm returns to Stage I.
\end{itemize}
\end{itemize}

The sampling algorithms for Stage I and Stage IIB are in section~\ref{kmc1}. For Stage IIA, we must determine how much local time the particle accumulates while it diffuses above the relevant patch. As in the case of the algorithm for the partially reactive patch, we can separately consider diffusion in the $\eta$ direction and the $(\mu, \nu)$ directions, allowing for greater computational efficiency. The algorithm first samples the time $t$ that it takes to diffuse to a circle of radius $d_1$ in the $(\mu, \nu)$ plane, where $d_1$ is the distance to the boundary. The algorithm samples a uniform random number $\xi \in [0,1]$ and numerically solves for $s^{*}$
\begin{align*}
P_c(s^{*}) = \xi,
\end{align*}
where $P_c(s)$ is given in \eqref{cyl_cdf}, which is then used to calculate the exit time $t = d_1^2 s^{*}$.

During this time $t$, the particle is guaranteed to experience the portion of the boundary contributing to the local time in the $\eta$ direction. Given the time $t$, by using the probability density function for the accumulated local time $\ell$ on the half-line \cite{grebenkov-surface-2020},
\begin{align*}
%p(\ell,t) = 
\frac{ \exp( - \ell^2/(4 t))}{\sqrt{\pi t}},
\end{align*}
we sample the accumulated local time using
\begin{align*}
\ell = 2 \sqrt{t} \, \text{erf}^{-1}(U),
\end{align*}
where $U$ is sampled uniformly on $[0,1]$. This local time $\ell$ is added to the total local time $L$. Then, the location of the particle in the $\eta$ direction is sampled based on the joint probability density function of $\eta$ and $\ell$ \cite{grebenkov-surface-2020}:
\begin{align*}
%\P(\eta, \ell, t) = 
\frac{\eta + \ell}{\sqrt{4 \pi t^3}} \exp \left( \frac{ -(\eta + \ell)^2}{4t} \right).
\end{align*}
This is done by sampling a uniform random variable $V$ on $[0,1]$ and computing
\begin{align*}
\eta = \sqrt{ \ell^2 - 4t \log(V)} - \ell.
\end{align*}
Finally, the algorithm updates the position of the particle as
\begin{align*}
(\mu, \nu, 0) + (d_1 \cos \theta, d_1 \sin \theta, \eta),
\end{align*}
where $\theta$ is sampled uniformly on $[ 0, 2 \pi]$ and $\eta$ was sampled above.

%%%%%%%%%%%%%%%%%%%%%%%%%%%%%%%%%
\begin{figure}%[b]
\centering
\includegraphics[width=.49\textwidth]{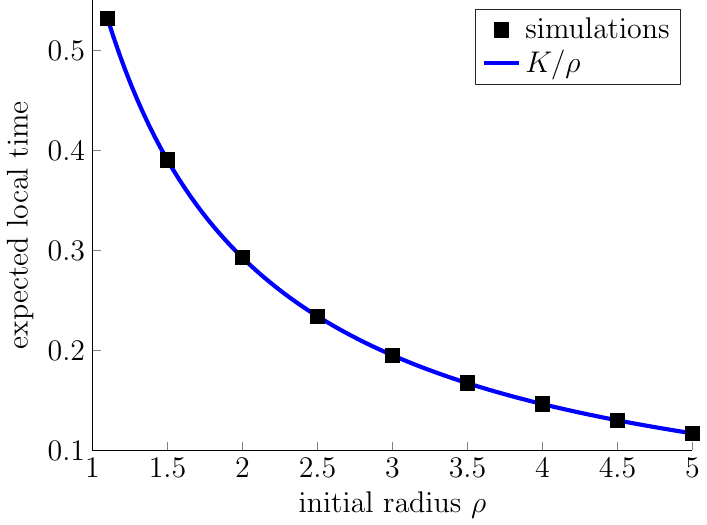}
\caption{Simulated expected local time in \eqref{writeout} as computed from this KMC simulation algorithm in Section~\ref{kmc2} (black square markers). The blue curve plots $K/\rho$ where $K\approx0.{5854}$ was chosen to fit this simulation data.}
\label{figlocal}
\end{figure}
%%%%%%%%%%%%%%%%%%%%%%%%%%%%%%%%%

%%%%%%%%%%%%%%%%%%%%%%%%%%%%%%%%%%%%%%%%%%%%%%%%%%%%%%%%%%%%%%%%%%%%%%%%%%%%%%%%%%%%%%%%%%
\subsection{Simulation results}

Figure~\ref{figlocal} plots the expected local time for a particle which begins uniformly distributed on the hemisphere of radius $\rho>0$,
\begin{align}\label{writeout}
\E[\ell_{\infty}\,|\,\mathbf{X}(0)\sim\text{uniform},\, \|\mathbf{X}(0)=\rho\|]
=-\overline{f}(\rho)
=\frac{K}{\rho},
\end{align}
as computed from this KMC simulation algorithm (black square markers). We ran this algorithm for starting radii $\rho \in \{1.1, 1.5, 2, \ldots, 4.5,5 \}$. For each value of $\rho$, we ran $10^6$ trials of the above algorithm with a maximum radius of $\rho_\infty = 10^{16}$ and estimated the average total accumulated local time $\overline{f}(\rho)$. The blue curve plots $K/\rho$ where $K\approx0.{5854}$ was chosen to fit this simulation data. In agreement with the theory, the maximum relative error between $K/\rho$ and the simulation data in Figure~\ref{figlocal} is less than $5\times10^{-4}$.

%%%%%%%%%%%%%%%%%%%%%%%%%%%%%%%%%%%%%%%%%%%%%%%%%%%%%%%%%%%%%%%%%%%%%%%%%%%%%%%%%%%%%%%%%%%%%%%%%%%%%%%%%%%%%%%%%%%%%%%%%%%%%%%%%%%%%%%%%%%%%%%%
\section{KMC for full stochastic system}\label{kmc3}

In this section, we describe the {{KMC}} algorithm developed to simulate the full stochastic system of a Brownian particle diffusing above a plane with a periodic square lattice of partially reactive patches (i.e.\ the patches are centered at $(x_{m},y_{n})=(m,n)$ for $m,n\in\mathbb{Z}$). We emphasize that the theoretical trapping rate formulas in Table~\ref{table_sum} were derived assuming only that the patches occupy a small fraction of the surface (see \eqref{mr}) and are well-separated (see \eqref{wellsep}), but we perform stochastic simulations in the special case that the patches are arranged on a square lattice. By sampling the times to absorption by any patch, we can numerically estimate the {trapping rate} $\chi$ of the plane by fitting the value of $\overline{\chi}$ in the cumulative distribution
\begin{align}
P(t, \overline{\chi}, z_0) = \text{erfc} \left( \frac{z_0}{2 \sqrt{t}} \right) - \exp \left( \overline{\chi}(\overline{\chi} t + z_0) \right) \text{erfc} \left( \frac{2 \overline{\chi} t + z_0}{2 \sqrt{t}} \right), \label{eq_cum_dist}
\end{align}
to the empirical cumulative distribution of the sampled times as described by Bernoff, Lindsay, and Schmidt \cite{bernoff-boundary-2018}. Equation~\eqref{eq_cum_dist} is the exact cumulative distribution function for the absorption time of a one-dimensional diffusion process with unit diffusivity that  starts distance $z_{0}\ge0$ from a partially absorbing boundary with reactivity $\overline{\chi}>0$ \cite{carslaw-conduction-1959}. This algorithm notably differs from the previous two algorithms in Sections~\ref{kmc1} and \ref{kmc2} in that the total running time $T$ is sampled and recorded, so it is necessary to sample the time for each stage.

%%%%%%%%%%%%%%%%%%%%%%%%%%%%%%%%%%%%%%%%%%%%%%%%%%%%%%%%%%%%%%%%%%%%%%%%%%%%%%%%%%%%%%%%%%%%%%%%%%%%%%%%%%%%%%%%%%%%%%%%%%%%%%%%%%%%
\subsection{Simulation algorithm}

Each simulation is initialized with a Brownian particle $\mathbf{Z}$ starting at position $(x_0, y_0, z_0)$, where $(x_0, y_0)$ are uniformly distributed on $[0,1]^2$ and $z_0$ is a fixed input parameter. The partially reactive patches have reactivity $\kappa$ and radii $\eps$, and are located on a $1 \times 1$ grid. The total running time $T$ is initialized at 0. The simulation proceeds by alternating between the following stages, similar to the previous two algorithms, allowing total time $T$ to accumulate until the particle is absorbed.
\begin{itemize}
\item \textit{Stage I: Project from bulk to plane:} The particle is projected to the $z = 0$ plane using the exact distribution, and the accumulated time is added to the total time $T$. If the particle lands on the plane such that it is within a disk, the algorithm proceeds to Stage IIA. If the particle lands on the plane outside of all disks, then the algorithm proceeds to Stage IIB.
\item \textit{Stage II: Project from plane to bulk:} Depending on where the particle landed, the particle either has a chance to be absorbed or does not before being reflected.
\begin{itemize}
\item \textit{Stage IIA: Within a Patch:} In the case that the particle lands on the plane within any patch, the algorithm computes the distance $d_1$ to the boundary of that patch and then samples the time $t$ until $(x,y)$ diffuses a distance $d_1$. Based on this time $t$, the algorithm samples whether or not the particle is absorbed. If the particle is absorbed, the time of absorption $\tau$ is sampled and added to the total time $T$, then the total time $T$ is recorded, and the trial ends. If the particle is not absorbed, the algorithm samples the particle's location in the bulk, adds the time $t$ to the total time $T$, and returns to Stage I.
\item \textit{Stage IIB: Outside All Patches:} The algorithm computes the distance $d_2$ to the boundary of the patch nearest. The particle is projected onto the hemisphere of radius $d_2$ following a uniform distribution. The time to diffuse that distance is sampled and added to the total time $T$, and the algorithm returns to Stage I.
\end{itemize}
\end{itemize}

The sampling algorithms for Stage I are in section~\ref{kmc1}. Before proceeding to Stage II, the algorithm adds the sampled time to diffuse to the plane $t^*$ to the cumulative time $T$.

The signed distance to the nearest patch at the start of Stage II given particle location $(x,y,0)$ and patch radii $\eps$ is calculated as
\begin{align*}
d = \sqrt{ ((x \text{ mod } 1) - 0.5)^2 + ((y \text{ mod } 1) - 0.5)^2 } - \eps.
\end{align*}
If $d \leq 0$, the particle landed within a patch, and the algorithm proceeds to Stage IIA with distance to boundary $d_1 = - d\ge0$. Otherwise, if $d > 0$, the particle landed outside all patches, and the algorithm proceeds to Stage IIB with distance to boundary $d_2 = d$.

In Stage IIA, we consider diffusion in the $z$ direction and the $(x,y)$ directions separately for efficiency, as in the previous two simulation algorithms. We first sample the time $t$ that it takes to diffuse to a circle of radius $d_1$ in the $(x,y)$ plane, and then we sample a uniform random number $\xi \in [0,1]$ and numerically solves for $\tau$
\begin{align*}
P_c(\tau) = \xi,
\end{align*}
where $P_c(\tau)$ is given in \eqref{cyl_cdf}. The algorithm then uses $\tau$ to calculate the exit time $t = d_1^2 \tau$. Given $t$, we sample the exit location of $z$ after time $t$ using the partial cumulative distribution $P_p(z, t, \kappa)$ given in \eqref{eq_absorb_loc}. Recall that this is a partial cumulative distribution, so the sampling algorithm may not return an exit location for $z$ indicating that the particle was absorbed, which occurs with the probability of being absorbed $Q(t,\kappa)$ given in \eqref{eq_p_ghost}. If the particle was absorbed (i.e., the algorithm does not return an exit location), then we need to sample the time to absorption $\tau^*$ conditioned that $\tau^{*} \leq t$. This conditional probability distribution for $s\in[0,t]$ is
\begin{align*}
\P(\tau^{*}\le s \,|\,\tau^{*}\le t)
=\frac{\P(\tau^{*}\le s)}{\P(\tau^{*}\le t)}
=\frac{1 - \exp \left( \kappa^2 s \right) \text{erfc} \big(\kappa \sqrt{s} \big)}{1 - \exp \left( \kappa^2 t \right) \text{erfc} \big(\kappa \sqrt{t} \big)}.
\end{align*}
To sample $\tau^{*}$ from this conditional distribution, we sample a uniform random variable $U$ on $[0,1]$ and numerically solve for the solution $\tau^*$ to the following equation,
\begin{align*}
U = \frac{1 - \exp \left( \kappa^2 \tau^{*} \right) \text{erfc} \big(\kappa \sqrt{\tau^{*}} \big)}{1 - \exp \left( \kappa^2 t \right) \text{erfc} \big(\kappa \sqrt{t} \big)}.
\end{align*}
The algorithm adds this time $\tau^*$ to the cumulative time $T$ and records $T$, and the trial ends.

If the particle was not absorbed, then the new position of the particle in the bulk is updated as
\begin{align*}
(x,y,0) + (d_1 \cos(\theta), d_2 \cos(\theta), z),
\end{align*}
where $\theta$ is a uniform random variable on $[0,2\pi]$ and $z$ was sampled above. The time $t$ for $(x,y)$ to diffuse a distance $d_1$ is added to the cumulative time $T$, and the simulation returns to Stage I.

In Stage IIB, the particle is propagated uniformly onto a hemisphere of radius $d_2$ centered at its current location $(x,y,0)$ according to
\begin{align*}
(x,y,0) + \frac{d_2}{\sqrt{ \xi_1^2 + \xi_2^2 + \xi_3^2}}( \xi_1, \xi_2, |\xi_3|),
\end{align*}
where each $\xi_i$ is a standard normal random variable. The time $t$ to diffuse this distance is sampled by numerically solving for $\tau'$:
\begin{align*}
P_S(\tau') = U,
\end{align*}
where $U$ is a uniform random variable on $[0,1]$ and $P_S(\tau')$ is the cumulative distribution for the exit time out of a hemisphere of radius $\pi$ \cite{bernoff-boundary-2018}. We precompute this cumulative distribution function using the short-time and long-time expansions given in \cite{bernoff-boundary-2018}:
\begin{align*}
P_S(\tau') = \begin{cases}
1 + 2 \sum_{n=1}^\infty (-1)^n e^{-n^2 \tau'} \quad & \text{(short-time expansion),}
\\
2 \sqrt{ \frac{\pi}{\tau'}} \sum_{n=0}^\infty \exp \left( - \pi^2 \left(n + \frac{1}{2} \right)^2 / \tau' \right) & \text{(long-time expansion).}
\end{cases}
\end{align*}
The time to diffuse the distance $d_2$ is given by $t = d_2^2 \tau' / \pi^2$, which is then added to the total cumulative time $T$, and the simulation returns to Stage I.

In order to fit the {trapping rate} $\overline{\chi}$ to a sample of times, we construct an empirical cumulative distribution by sorting the sample of times from least to greatest, so $t_1 < t_2 < \cdots < t_N$, and define the empirical cumulative distribution as
\begin{align*}
P_e(t_j) = \frac{j-\frac{1}{2}}{N},
\end{align*}
which we can then compare to the homogenized cumulative distribution $P(t, \overline{\chi},z_{0})$ given in \eqref{eq_cum_dist} \cite{bernoff-boundary-2018}. We search for the value of $\overline{\chi}$ that minimizes the Kolmogorov-Smirnov distance \cite{plunkett-boundary-2023}
\begin{align*}
\mathcal{E}(\overline{\chi}) = \max_{t \in \{t_1, \ldots, t_N \} } \left| P(t, \overline{\chi},z_{0}) - P_e(t) \right|.
\end{align*}

%%%%%%%%%%%%%%%%%%%%%%%%%%%%%%%%%
\begin{figure}%[tb]
\centering
(a)\hspace{-.05\textwidth}\includegraphics[width=.47\textwidth]{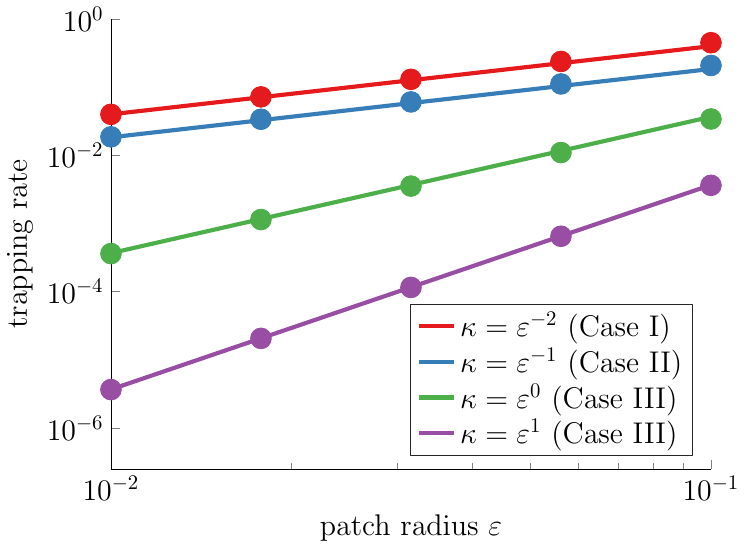}
\hspace{.05\textwidth}
(b)\hspace{-.05\textwidth}\includegraphics[width=.47\textwidth]{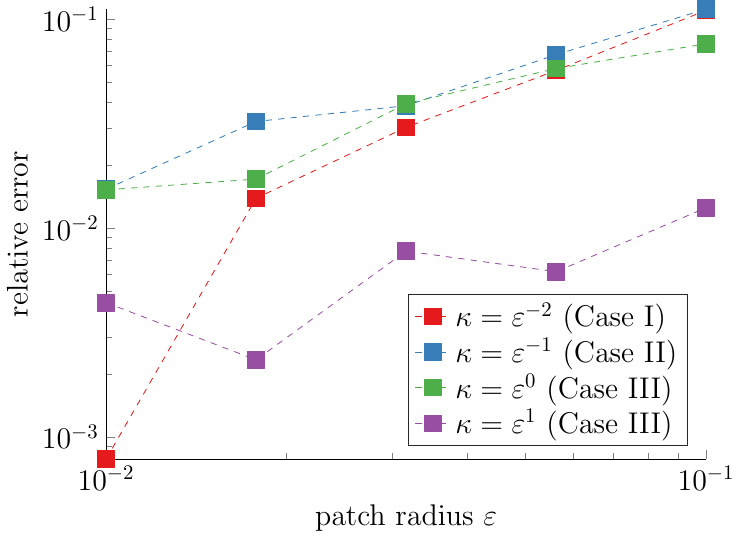}
\caption{Comparison of simulations of the full stochastic absorption process to the theoretical trapping rates. See the text for details.}
\label{figfull}
\end{figure}
%%%%%%%%%%%%%%%%%%%%%%%%%%%%%%%%%

%%%%%%%%%%%%%%%%%%%%%%%%%%%%%%%%%%%%%%%%%%%%%%%%%%%%%%%%%%%%%%%%%%%%%%%%%%%%%%%%%%%%%%%%%%%%%%%%%%%%%%%%%%%%%%%%%%%%%%%%%%%%%%%%%%%%%%%%%%%%%%%%%%%%%%%%%%%%%%%%%%%%%
\subsection{Simulation results}

In Figure~\ref{figfull}, we compare the results of the simulation algorithm above compared to our theoretical trapping rates. In Figure~\ref{figfull}a, we show results (circle markers) for trapping rates $\kappa$ which are given by $\kappa=\eps^{-2}$, $\kappa=\eps^{-1}$, $\kappa=\eps^{0}=1$, and $\kappa=\eps^{1}$ (simulation results are for $10^{5}$ trials for each value of $\eps$ and $\kappa$). In this plot, $\kappa=\eps^{-2}$ is compared to to the Case I theory in \eqref{eq_reacs}, $\kappa=\eps^{-1}$ is compared to to the Case II theory in \eqref{eq_reacs}, and $\kappa=\eps^{0}=1$ and $\kappa=\eps^{1}$ are compared to the Case III theory in \eqref{eq_reacs} (the curves are the theoretical values, where $c_{0}(1)$ and $K$ are computed from the KMC simulations in Sections~\ref{kmc1}-\ref{kmc2}). Figure~\ref{figfull}b plots the relative errors between the simulation results and the theoretical trapping rates.

%%%%%%%%%%%%%%%%%%%%%%%%%%%%%%%%%%%%%%%%%%%%%%%%%%%%%%%%%%%%%%%%%%%%%%%%%%%%%%%%%%%%%%%%%%%%%%%%%%%%%%%%%%%%%%%%%%%%%%%%%%%%%%%%%%%%%%%%%%%%%%%%%%%%%%%%%%%%%%%%%%%%%%%%%%%%%%%%%%%%%%%%%%%%%%%%%%%%%%
\section{Comparison to prior estimates}\label{comparison}

As described in the Introduction, the problem of homogenizing a patchy boundary consisting of partially reactive patches with reactivity $\kappa_{\text{p}}$ has been addressed previously via heuristic means. In particular, the following formula has been posited for the trapping rate for this scenario \cite{zwanzig-time-1991, berezhkovskii-boundary-2004},
\begin{align}\label{interp2}
\overline{\kappa}_{\text{heur}}
=\frac{\sigma\kappa_{\text{p}}\overline{\kappa}_{0}}{\sigma\kappa_{\text{p}}+\overline{\kappa}_{0}},
\end{align}
where $\overline{\kappa}_{0}$ is the trapping rate for the corresponding problem involving perfectly reactive patches and $\sigma\in(0,1)$ denotes the fraction of the surface covered by patches. If $\sigma\ll1$ and the patches are disks with common radius $a>0$, then $\overline{\kappa}_{0}=4D\sigma/(\pi a)$ \cite{berg-physics-1977} and \eqref{interp2} becomes
\begin{align}\label{interp3}
\overline{\kappa}_{\text{heur}}
=\frac{4D\sigma}{\pi a} \frac{a\kappa_{\text{p}}/D}{a\kappa_{\text{p}}/D+4/\pi}.
\end{align}
To our knowledge, the heuristic formulas in \eqref{interp2}-\eqref{interp3} have not been systematically derived. Instead, \eqref{interp3} has been justified by noting that (i) \eqref{interp3} has the desired limiting behavior of $\overline{\kappa}_{\text{heur}}\to4D\sigma/(\pi a)$ if $a\kappa_{\text{p}}/D\to\infty$ and $\overline{\kappa}_{\text{heur}}\to0$ if $a\kappa_{\text{p}}/D\to0$ and (ii) \eqref{interp3} has shown agreement with stochastic simulations \cite{zwanzig-time-1991, berezhkovskii-boundary-2004}.

We now compare our trapping rate (in dimensional units),
\begin{align}\label{main2}
\overline{\kappa}
\approx
\begin{dcases*}
\frac{4D\sigma}{\pi a} & \text{if }$a\kappa_{\text{p}}/D\gg1$,\\
\frac{4D\sigma}{\pi a}\frac{c_{0}(a\kappa_{\text{p}}/D)}{2/\pi} & \text{if }$a\kappa_{\text{p}}/D=O(1)$,\\
\frac{4D\sigma}{\pi a}\frac{a\kappa_{\text{p}}/D}{2/\pi}K & \text{if }$a\kappa_{\text{p}}/D\ll1$.
\end{dcases*}
\end{align}
to the previously suggested trapping rate in \eqref{interp3}. The trapping rates in \eqref{interp3} and \eqref{main2} clearly agree in the limit $a\kappa_{\text{p}}/D\to\infty$. For $a\kappa_{\text{p}}/D=O(1)$, the agreement between \eqref{interp} and \eqref{main2} is equivalent to the agreement of the following sigmoidal approximation,
\begin{align}\label{sapprox}
c_{0}(a\kappa_{\text{p}}/D)
\approx\frac{(2/\pi)a\kappa_{\text{p}}/D}{a\kappa_{\text{p}}/D+4/\pi}.
\end{align}
As shown in Section~\ref{sigmoid}, the sigmoidal approximation in \eqref{sapprox} turns out to be generally accurate, with the greatest discrepancies arising when $a\kappa_{\text{p}}/D$ is small (see Figure~\ref{figc0}). Indeed, the relative error in \eqref{sapprox} rises above 10\% for $a\kappa_{\text{p}}/D<1/3$. Consistent with this point, the greatest discrepancy between the trapping rates \eqref{interp3} and \eqref{main2} arises in the limit $a\kappa_{\text{p}}/D\to0$, 
\begin{align*}
\frac{\overline{\kappa}-\overline{\kappa}_{\text{heur}}}{\overline{\kappa}_{\text{heur}}}
\to\frac{K-1/2}{1/2}
\approx17\%\quad\text{as }a\kappa_{\text{p}}/D\to0,
\end{align*}
where we have used our numerically computed value $K\approx0.{5854}$ from Section~\eqref{kmc2}.

To summarize, the heuristic trapping rate formula in \eqref{interp3} generally shows a good agreement with our trapping rate in \eqref{main2} which was derived by combining asymptotic analysis and numerical computation. The trapping rates agree in the (trivial) limit of $a\kappa_{\text{p}}/D\gg1$, and the relative difference between the trapping rates approaches about 17\% for $a\kappa_{\text{p}}/D\ll1$.

%%%%%%%%%%%%%%%%%%%%%%%%%%%%%%%%%%%%%%%%%%%%%%%%%%%%%%%%%%%%%%%%%%%%%%%%%%%%%%%%%%%%%%%%%%%%%%%%%%%%%%%%%%%%%%%%%%%%%%%%%%%%%%%%%%%%%%%%%%%%%%%%%%%%%%%%%%%%
\section{Discussion}\label{discussion}

In this paper, we performed boundary homogenization to derive the trapping rate for a surface containing partially reactive patches. We first formulated the problem in terms of a stochastic process and an associated PDE boundary value problem. In the limit that the patches occupy a small fraction of the surface, we used matched asymptotic analysis, double perturbation expansions, and homogenization theory to derive three formulas for the trapping rate that apply in different cases of relative sizes of the patch radii to their reactivities. Using results from probability theory, we developed two {{KMC}} simulation algorithms to calculate factors appearing in these trapping rates. We developed a third {{KMC}} simulation algorithm to simulate the full stochastic absorption process which confirmed our theoretical trapping rate formulas.

As described in the Introduction, the problem of homogenizing a surface with partially reactive patches has been previously studied using a heuristic formula which interpolates between the case that the patches are perfectly reactive and the case that the patches are perfectly reflective \cite{zwanzig-time-1991, berezhkovskii-boundary-2004}. We showed that this heuristic formula is generally quite accurate when the patches occupy a small fraction of the surface, with the largest error of around 17\% occurring for patches with low reactivity.

Prior work on boundary homogenization has primarily considered perfectly reactive patches \cite{muratov-boundary-2008, bernoff-boundary-2018, berezhkovskii-homogenization-2006, belyaev-effective-1999, berezhkovskii-boundary-2004, makhnovskii-trapping-2006, plunkett-boundary-2023, zwanzig-diffusion-controlled-1990, dagdug-boundary-2016, lindsay-first-2017, lawley-how-2019, plunkett-bimolecular-2021, lawley-boundary-2019}. Recently, some other groups have considered diffusive interactions with partially reactive patches and targets \cite{schumm-search-2021, bressloff-narrow-2022, chaigneau-first-passage-2022}. To our knowledge, the {{KMC}} simulation algorithms for absorption at partially reactive patches and local time accumulation are the first such {{KMC}} simulation algorithms. 

For simplicity, we assumed that the partially reactive patches are disks. However, our analysis could be extended to more general patch shapes. Indeed, our simulation algorithms can be immediately extended to arbitrary patch shapes on a flat surface, assuming that the distance between any point on the surface and the patch boundary can be efficiently computed. 

%%%%%%%%%%%%%%%%%%%%%%%%%%%%%%%%%%%%%%%%%%%%%%%%%%%%%%%%%%%%%%%%%%%%%%%%%%%%%%%%%%%%%%%%%%%%%%%%%%%%%%%%%%%%%%%%%%%%%%%%%%%%%%%%%%%%%%%%%%%%%%%%%%%%%%%%%%%

\bibliography{robin_library}
\bibliographystyle{siam}

\end{document}